\DocumentMetadata{}
\documentclass[nonacm, screen]{acmart}
\usepackage{amsfonts}
\usepackage{amsmath}
\usepackage{bm}
\newcommand{\ind}{\perp\!\!\!\!\perp} 
\newcommand{\dep}{\not \! \perp \!\!\!\! \perp}
\newtheorem{definition}{Definition}
\newtheorem{assumption}{Assumption}
\usepackage{multirow}
\usepackage{tabularx}
\usepackage{color}
\usepackage{threeparttable}
\usepackage{subfig}
\usepackage{enumitem}
\usepackage{rotating}
\usepackage{makecell}
\usepackage{array}
\usepackage{booktabs}

\AtBeginDocument{%
  \providecommand\BibTeX{{%
    \normalfont B\kern-0.5em{\scshape i\kern-0.25em b}\kern-0.8em\TeX}}}

\begin{document}

\title{A Survey on Causal Inference for Recommendation}

\author{Huishi Luo}
\email{hsluo2000@buaa.edu.cn}
\orcid{0000-0002-3553-2280}
\author{Fuzhen Zhuang}
\authornote{Corresponding author: zhuangfuzhen@buaa.edu.cn.}
\email{zhuangfuzhen@buaa.edu.cn}
\orcid{0000-0001-9170-7009}
\affiliation{%
  \institution{Institute of Artificial Intelligence, Beihang University}
  \city{Beijing}
  \country{China}
  \postcode{100191}
}

\author{Ruobing Xie}
\email{ruobingxie@tencent.com}
\orcid{0000-0003-3170-5647}
\affiliation{%
  \institution{WeChat Search Application Department, Tencent}
  \city{Beijing}
  \country{China}
  \postcode{100080}
}

\author{Hengshu Zhu}
\email{zhuhengshu@gmail.com}
\orcid{0000-0003-4570-643X}
\affiliation{%
  \institution{the Career Science Laboratory, BOSS Zhipin}
  \city{Beijing}
  \country{China}
  \postcode{100028}
}

\author{Deqing Wang}
\email{dqwang@buaa.edu.cn}
\orcid{0000-0001-6441-4390}
\affiliation{%
  \institution{SKLSDE, School of Computer Science, Beihang University}
  \city{Beijing}
  \country{China}
  \postcode{100191}
}

\author{Zhulin An}
\email{anzhulin@ict.ac.cn}
\author{Yongjun Xu}
\email{xyj@ict.ac.cn}
\affiliation{%
  \institution{Institute of Computing Technology, Chinese Academy of Sciences}
  \city{Beijing}
  \country{China}
}

\renewcommand{\shortauthors}{Luo et al.}

\begin{abstract}
  Causal inference has recently garnered significant interest among recommender system (RS) researchers due to its ability to dissect cause-and-effect relationships and its broad applicability across multiple fields. It offers a framework to model the causality in recommender systems like confounding effects and deal with counterfactual problems such as offline policy evaluation and data augmentation. Although there are already some valuable surveys on causal recommendations, they typically classify approaches based on the practical issues faced in RS, a classification that may disperse and fragment the unified causal theories. Considering RS researchers' unfamiliarity with causality, it is necessary yet challenging to comprehensively review relevant studies from a coherent causal theoretical perspective, thereby facilitating a deeper integration of causal inference in RS. This survey provides a systematic review of up-to-date papers in this area from a causal theory standpoint and traces the evolutionary development of RS methods within the same causal strategy. Firstly, we introduce the fundamental concepts of causal inference as the basis of the following review. Subsequently, we propose a novel theory-driven taxonomy, categorizing existing methods based on the causal theory employed—namely, those based on the potential outcome framework, the structural causal model, and general counterfactuals. The review then delves into the technical details of how existing methods apply causal inference to address particular recommender issues. Finally, we highlight some promising directions for future research in this field. Representative papers and open-source resources will be progressively available at \url{https://github.com/Chrissie-Law/Causal-Inference-for-Recommendation}.
\end{abstract}




\maketitle

\section{Introduction}
\label{sec:intro}
Recommender systems (RS), working as filtering systems to present personalized information to users and alleviate information overload, have been widely deployed in various online applications, including e-commerce, social networks, and multimedia services. Recently, an emerging research direction has attracted increasing attention from RS researchers, which explores the integration of advanced machine learning with a traditional statistics field, causal inference. Causal inference~\cite{gelman2011causality, imbens2015causal} works to analyze the relationship between a cause and its effect~\cite{pearl2009causality}, which has a wide range of real-world applications in both academic and industrial domains, such as medicine~\cite{kessler2019machine, shalit2020can}, climate~\cite{tan2021associations}, political science~\cite{schlotter2011econometric}, and online advertising evaluation~\cite{li2016matching, fong2018covariate}. Treatment effect estimation is a fundamental problem in causal inference, often applied in policy evaluation. For example, in pharmaceutical research, where we are interested in the effect of a drug on lifespan, we need to answer a causal question involving a so-called intervention or treatment: What is the probability that a typical patient would survive in $L$ years if made to take the drug? A large-scale randomized controlled trial is the golden solution but it may suffer from the expense and even ethical issues. Therefore, in most cases, we can only estimate the effect from non-randomized observational data, where the correlation between the drug and survival does not imply causation, because factors including age, gender, and severity of the disease may affect the outcomes. 

\begin{figure}[t!]
\centering
\vspace{-1mm}
\includegraphics[height=0.28\textheight,trim=120 0 100 0 ,clip]{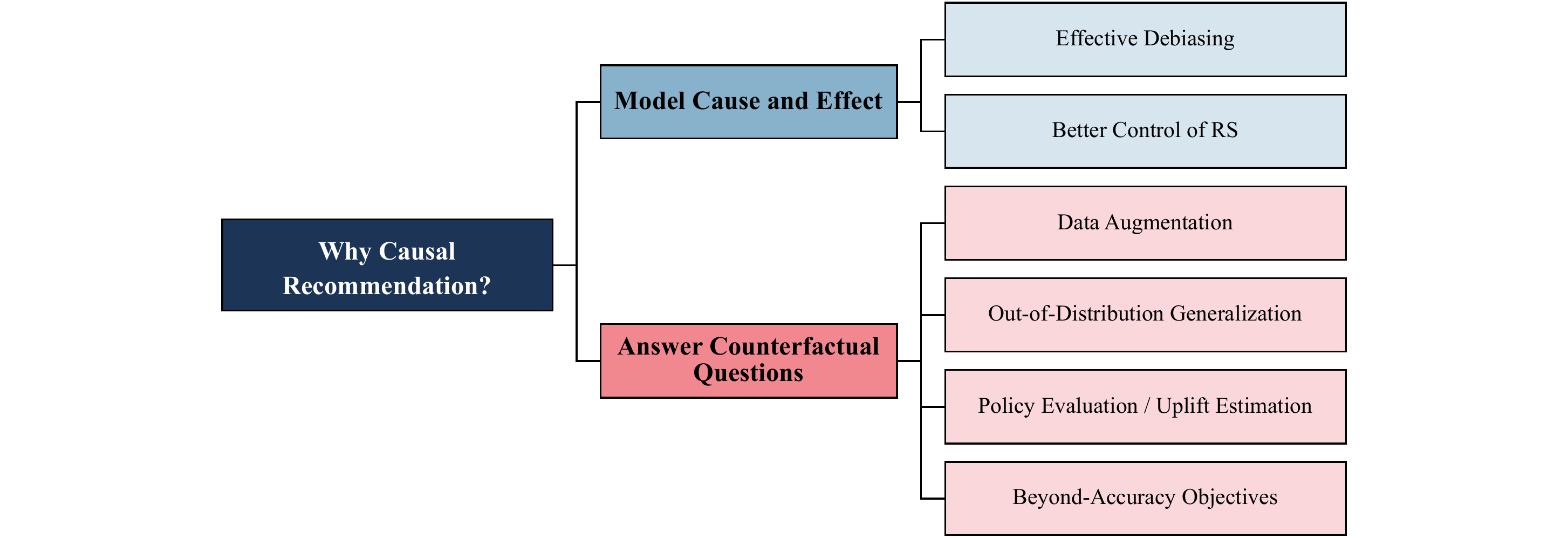} 
\vspace{-1mm}
\caption{Strengths of causal inference for recommendation.}
\vspace{-6mm}
\label{fig:strength}
\end{figure}

Causality for recommendation has been widely used in uplift modeling for policy effect evaluation~\cite{radcliffe2007using, gutierrez2017causal}, but it was not until the last few years that research has tended to focus on applying it to model training. Common recommendation scenarios in practice, including click-through rate (CTR) prediction and post-click metric prediction, etc., can be abstracted into causal problems, and causal inference can be applied in different stages of the entire RS project, such as preliminary data collection~\cite{ wang2021counterfactual}, representations learning of users and items embeddings~\cite{ liu2021mitigating, he2022causpref, wang2022causal}, objective optimization~\cite{ mehrotra2018towards, mcinerney2020counterfactual, bonner2018causal}, and policy evaluation offline and online~\cite{ sato2019uplift, saito2021counterfactual, sato2021online}. Causal recommender systems can surpass traditional approaches, primarily due to two key strengths (Fig. \ref{fig:strength}):

1. \textbf{Model cause and effect.} The majority of current machine learning systems, including RS, operate predominantly in a statistical mode~\cite{pearl2018theoretical, xu2023artificial}, which focuses on the correlation between variables. However, in applications, we care more about causality rather than correlation, and it is well known that “Correlation is not causation”. For example, a movie recommendation platform recorded a female user who has finished watching an action movie, so it concludes that she likes action movies and makes many recommendations for related action movies. Nevertheless, the user may have watched the movie due to its popularity rather than her inherent preference for action movies in fact. Therefore, the spurious correlation between user interest and movie genres learned by traditional recommender systems may lead to a degraded user experience. In contrast, causal recommendation systems can learn the causal effects of users' individual interest as well as conformity on the interaction outcome (i.e., watching), respectively, so that action movies will not be incorrectly recommended later. Modeling cause and effect enables causality-based recommender systems to 1) measure the causal effects on user interaction of source variables of a wide range of bias, such as popularity~\cite{zhang2021causal, wei2021model} and exposure~\cite{liang2016modeling, wang2021clicks}, thus performing effective debiasing, which is currently the most common application of causal inference for recommendation; 2) better control of RS due to decomposition and inference of the causal effect of variables, for example, leveraging the causal effect of certain bias to improve recommendation accuracy~\cite{zhang2021causal}.

2. \textbf{Answer counterfactual questions.} Many recommender system problems, including data augmentation, out-of-distribution (OOD) generalization, and policy evaluation, are essentially counterfactual problems, that is, the situation where the values of some causal variables are different from reality. 1) In terms of the data augment problem, as a significant complementary resource of the observed data~\cite{ wang2021counterfactual}, the counterfactual data needs to answer questions such as "What would be the user’s interaction if the recommended items had been different?" or "What would be the probability of click if an item had been recommended to a user who has not been recommended before?”. 2) The OOD problem refers to the recommendation which violates the Independent and Identically Distributed (IID) assumption of the interactions between training and testing periods~\cite{he2022causpref}. Traditional recommendation may learn false associations between users and items, while causal recommender systems adopt the counterfactual means to find invariant or unchangeable variables or causal relationships in the recommendation task and reuse them to generalize when the distribution changes. For example, if a pregnant female user had purchased red high heels before pregnancy, a traditional recommendation system might continue to make recommendations of high heels, but a causal inference system can learn the causal relationship between high heels and pregnancy status through causal tools like causal graph. Therefore, when the user's status shifts (identified from the user's behavior like purchasing baby products), the causal recommender system no longer recommends high-heeled shoes but retains the user's preference for the red color and recommends red clothing instead. 3) Uplift modeling estimates the increase, or uplift, in user interactions caused by recommendations, which is a counterfactual problem because one needs to estimate the difference between two mutually exclusive outcomes for an item (either item $i$ is recommender or not for a specific user). 4) In addition to the above issues, counterfactuals can also optimize beyond-accuracy objectives such as fairness and explainability. For example, to ensure fairness, sensitive features like sex and race can be modified and removed in the counterfactual world~\cite{li2021towards}, and explainability can be attained by comparing the real world with the counterfactual world to search for user interactions that affect recommendation results\cite{ ghazimatin2020prince, tan2021counterfactual}.

It is worth mentioning that there are some existing surveys~\cite {wu2022opportunity, gao2022causal,zhu2023causal, xu2023causal} of causal inference for recommender systems. However, the present study distinguishes itself from these previous works for several reasons.  

\begin{figure*}[t!]
\centering
\vspace{-3mm}
\includegraphics[height=0.56\textheight,trim=100 0 0 0 ,clip]{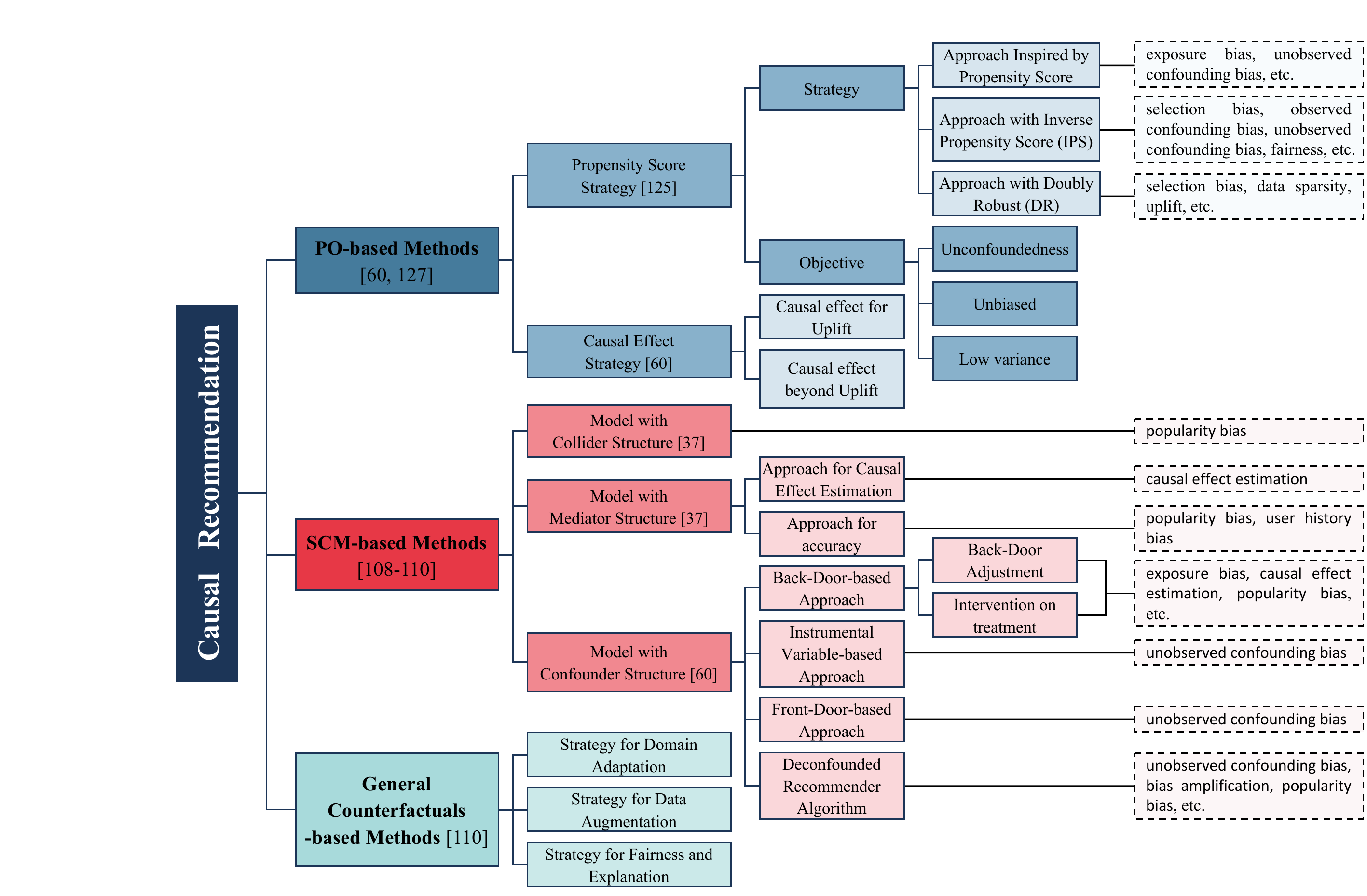} 
\vspace{-2mm}
\caption{Strategies of the causal inference for recommendation.}
\label{fig:classification}
\end{figure*}

        1. \textbf{Theoretically coherent classification framework from a causal perspective.} The aforementioned surveys fall short in providing a comprehensive taxonomy of causal recommender systems. Specifically, ~\cite{wu2022opportunity} only discusses recommendation methods of potential outcome framework~\cite{rubin1974estimating}, and approaches investigated in ~\cite{gao2022causal,zhu2023causal,xu2023causal} are mainly classified from the application perspectives, i.e., issues of recommender systems. This application-centric taxonomy, while practical, tends to obscure the underlying theoretical coherence of causal inference methods, as a single causal theory could be applied in various problems. Contrastingly, our survey adopts a more nuanced and theory-driven classification, involving the Neyman-Rubin potential outcome (PO) framework ~\cite{ splawa1990application, rubin1974estimating} and the Pearl structural causal model (SCM) framework~\cite{pearl1995causal, pearl1988probabilistic, pearl2009causality}. Within this paper, causal-based recommendation algorithms are categorized into three main types: PO-based, SCM-based, and general counterfactuals-based. Both PO-based and SCM-based methods utilize specific causal inference techniques, but the former does not explicitly employ causal structure information. On the other hand, general counterfactuals-based methods refer to those designed under the inspiration of counterfactual concepts, without using particular causal inference techniques. The classification framework is illustrated in Fig. \ref{fig:classification}. This taxonomy not only provides a more structured and holistic understanding of causal theories but also empowers researchers, particularly newcomers to the field of causal inference, to effectively grasp and apply these theories in practice.

2. \textbf{Evolution of Causal Methods in Recommender Systems.} We systematically delineates the developmental trajectory of the integration between prevalent causal inference theories and recommender systems, as illustrated in Fig. \ref{fig:propensity} and \ref{fig:he_type}. Through this intuitive exposition, readers can readily perceive how methodologies within a specific domain have been iteratively proposed and the particular issues they address in their respective evolutions.

\begin{figure}[t!]
\centering
\vspace{-2mm}
\includegraphics[height=0.22\textheight,trim=0 0 0 0 ,clip]{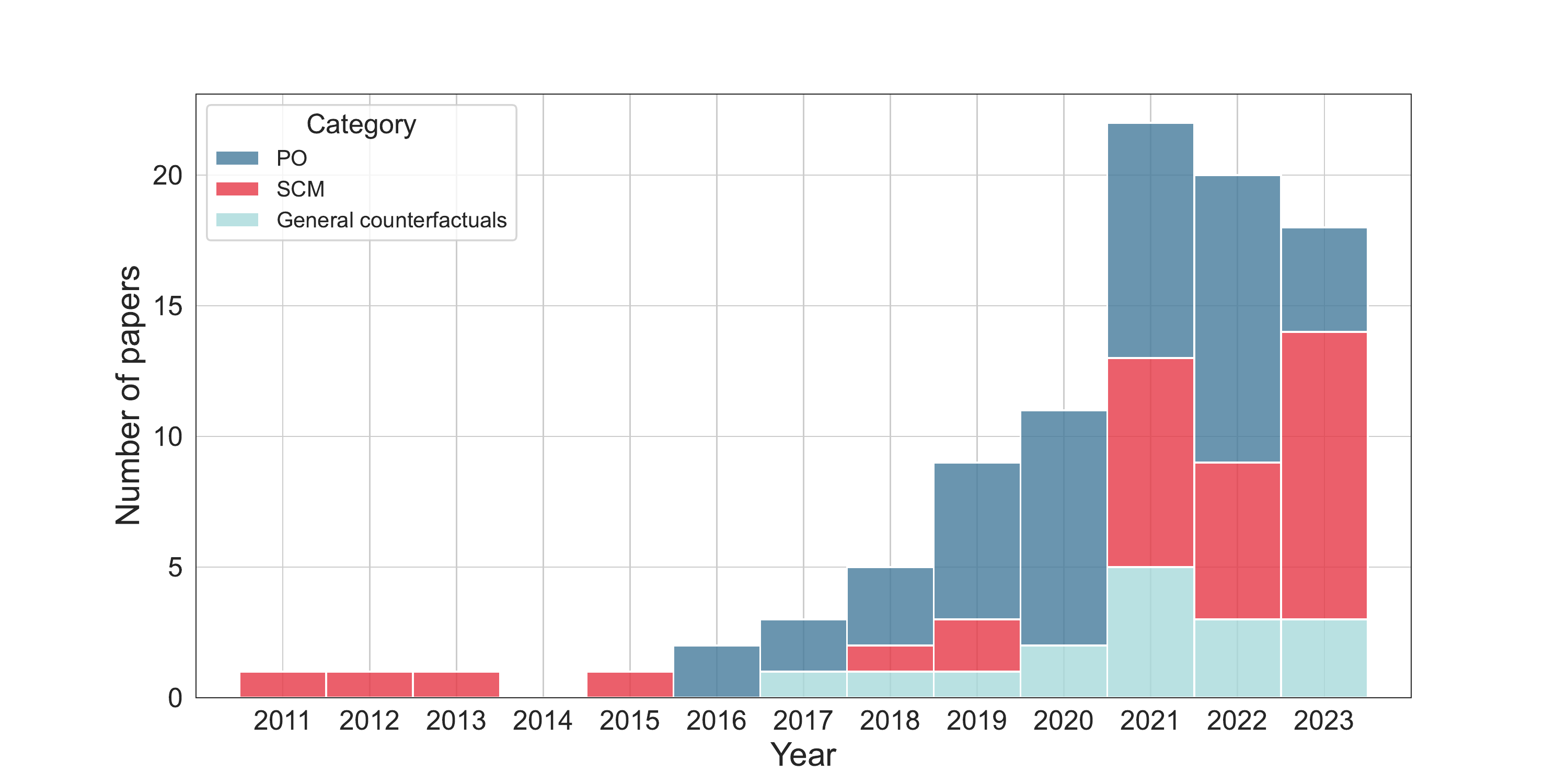} 
\vspace{-4mm}
\caption{Distribution of publications on causal recommendations by year and framework, focusing exclusively on specific industrial algorithms and excluding fundamental theory discussions.}
\vspace{-3mm}
\label{fig:statistics}
\end{figure}

3. \textbf{Up-to-Date Collection and Review.} Given the growing popularity of this domain, our survey encompasses numerous recent publications absent in~\cite{wu2022opportunity, gao2022causal,zhu2023causal,xu2023causal}. We have collected papers related to causal inference-based recommender algorithms from esteemed conference proceedings and journals, and visualize the statistics of them concerning the published year and causal inference framework in Fig. \ref{fig:statistics}.

This paper provides a comprehensive summary of the work on causal recommender systems. The rest of this survey is organized into six sections. 
Section \ref{sec:foundation} introduces the basic concepts within recommendation and causal inference of both frameworks, and highlights the distinctions between this survey and existing reviews in this field. Sections \ref{sec:po_based}, \ref{sec:scm_based}, and \ref{sec:general_counterfactual} interpret causal recommendation approaches from the perspective of causal techniques: PO-based, SCM-based, and general counterfactuals-based, respectively. Future research directions are openly discussed in Section \ref{sec:discuss}. The last section concludes this survey. Detailed discussions on related fields and more foundational concepts of causal inference theories, are presented in the Supplemental Information. The main contributions of this survey are summarized below.

\begin{itemize}
    \item Novel taxonomy: We separate various causal recommendation methods into three major categories based on the causal framework they adopt, which may be more instructive than existing taxonomies for the readers to integrate causal inference with recommender systems and propose new approaches in practice (see Section \ref{sec:exist_cate}).

    \item Comprehensive review: Over one hundred and twenty causal recommendation papers from the last century to 2023 are introduced, explained, and summarized, which might give readers a comprehensive overview of causality for recommendation.

    \item Open discussion: Research directions for applying causality to improve recommendation methods in the academic and industrial areas are openly discussed. 
\end{itemize}

\section{Related fields}
\label{sec:related}
Some areas related to causal inference may be unfamiliar to recommender system researchers; thus, we introduce them and carefully clarify the connections and differences between them and causal inference.

\textbf{Causal Discovery~\cite{ peters2017elements}:} Causal discovery is a crucial technique in causality. Causality is the science of cause and effect~\cite{ pearl2018book}. In Pearl’s theory, causality contains two fundamental problems: the first is to prove that one variable is the cause of another or find the cause of a variable, and the second is to draw a conclusion of what might be the effects if changing the value of a variable. The former corresponds to causal discovery, also called causal structure learning,  seeking to discover causal relations, which are stable physical mechanisms in nature and manifest themselves in determined functional relationships between variables, from the data based on some causal assumptions that are hardly testable in observational studies~\cite{pearl2010causal}. The latter corresponds to causal inference, which estimates the outcome after an intervention with a given causal relationship (usually a causal structure obtained by causal discovery or empirically based hypotheses) ~\cite{pearl2018book, yao2021survey, pearl2009causality}. In other words, the former is the basis of the latter, because it is impossible to tell what variables would be affected by an intervention without a causal structure, and thus no interventions and counterfactuals can be implemented. For example, we cannot determine the effect of opening umbrellas on rain if we do not know the causal relevance between them since they always coincide. Note that most methods for causal discovery rely on the SCM framework~\cite{naser2022causality}.

\textbf{Bayesian Inference:} Bayesian inference is a popular approach to data analysis based on Bayes’ theorem, where all observed and unobserved parameters are given a joint probability distribution, i.e., the prior and data distributions, to inference prior distribution~\cite{ van2021bayesian}. Bayesian inference is regarded as one of key techniques and integral components in both PO and SCM: In PO, with assignment mechanisms and the definition of potential outcomes, a Bayesian model can be used to connect the treatment and potential outcomes in real world or counterfactual world; in SCM, Bayesian networks are widely used as causal graphs to present causal associations between variables. Nevertheless, there is a primary distinction between Bayesian inference and causal inference. Bayesian inference is causality-free statistics that focus on associations, such as dependence, likelihood, etc., which can be formulated in terms of distribution functions. However, what is unique to causal inference is that causal concepts cannot be defined from statistics associations alone. As the example mentioned above, it is impossible to tell from the statistics whether raining causes the behavior of opening umbrellas or vice versa. This core distinction leads to two differences between Bayesian inference and causal inference in their specific manifestations, including assumptions and notations. 1) Bayesian inference is based on associational assumptions, which, even untested, are testable in principle~\cite{pearl2010causal}. However, as for causal inference, causal assumptions, in contrast, cannot be verified even in principle unless we proactively influence the observed data, i.e., resort to experimental control. In general, the sensitivity to priors in Bayesian statistics, such as the IID assumption, will decrease with increasing sample size, while sensitivity to prior causal assumptions, say that whether to open umbrellas does not affect the weather, remains substantial regardless of sample size. 2) New notations are introduced to causal inference as causal expressions compared with Bayesian statistics, which is presented in detail in Section ~\ref{sec:foundation}.
\section{Foundation}
\label{sec:foundation}

In this section, background information and several important concepts of causal inference and recommender systems are introduced to facilitate readers' understanding of the inter-study of the two research fields. The notations used in this survey are listed for convenience. At the end of this section, we set up categorizations of causal recommendations.

\subsection{Causal Inference}
In this part, we will give a brief review of two representative frameworks of causal inference, including the potential outcome (PO) framework by Rubin et al.~\cite{splawa1990application, rubin1974estimating, imbens2015causal} and the structural causal models (SCM) framework by Pearl et al.~\cite{pearl1995causal, pearl1988probabilistic, pearl2009causality}
. Note that these two frameworks are logically equivalent~\cite{ pearl2009causality}.

\subsubsection{Potential Outcome Framework}
\label{sec:po}

The Potential Outcomes Framework (aka the Neyman-Rubin Causal Model) ~\cite{splawa1990application, rubin1974estimating, imbens2015causal} is the most widely used framework across many disciplines. With a hypothetical treatment (or manipulation, intervention), the causal effect, i.e., treatment effect, is defined as the difference between the potential outcomes under treatment and control for \textit{the same unit}~\cite{imbens2015causal}.

\begin{definition}[Unit]
A unit refers to the research object in the potential framework.
\end{definition}

A unit can be a physical object, an individual, or a collection of objects or persons, such as a classroom or a market, at a particular point in time~\cite{imbens2015causal}. In recommendation research, a user-item pair will usually be defined as a unit. It should be noticed that the same physical object or person at a different time is a different unit. This is a reasonable restriction, considering the same user will make different decisions at a different time even if exposed to the same item due to factors like preference shift, mood, occasion and so on.

\begin{definition}[Treatment]
Treatment can be defined as the action applied to a unit.
\end{definition}

This paper focuses on binary treatment (e.g., recommend or not), the most common setting in the recommendation field. In practice, we refer to the more active treatment simply as the “treatment” $T=1$ and the other treatment as the “control” $T=0$.

\textbf{Potential Outcome.} For each treatment-unit pair, the potential outcome is the outcome that the treatment is applied to the unit, denoted as $Y(T=t)$ (ignoring unit). For a unit, only the potential outcome corresponding to the treatment actually taken will be observed, denominated as observed outcome, while others are referred to as counterfactual outcomes. The \textit{fundamental problem of causal inference} in PO framework is that we can never obtain both observed and counterfactual outcomes for a unit: it is impossible to realize all treatments and observe the corresponding outcomes.

\textbf{Treatment Effect/ Causal effect.} Treatment effect is represented by the difference between the potential outcomes under treatment and control for the same unit, formulated as:

\begin{equation}
\label{eq:ITE}
    \textrm{TE} = Y(T=1) - Y(T=0),
\end{equation}

where $ {Y}(T=1)$ and $ {Y}(T=0)$ are the potential treated and control outcome of the unit, respectively. Treatment effect like Equation \ref{eq:ITE} is also called \textbf{Individual Treatment Effect}. Furthermore, the treatment effect can be defined at the population and subpopulation levels. At the population level, \textbf{ Average Treatment Effect (ATE) } is the expectation of ITE over the whole population~\cite{guo2020survey}, denoted as:

\begin{equation}
\label{eq:ATE}
    \textrm{ATE} = \mathbb{E}[Y(T=1) - Y(T=0)].
\end{equation}

The ATE on the subpopulation level is often of particular interest; thus we define \textbf{Conditional Average Treatment Effect (CATE)} on the units with the same features $X=x$ as:
\begin{equation}
\label{eq:CATE}
    \textrm{CATE} = \mathbb{E}[Y(T=1|X=x) - Y(T=0|X=x)].
\end{equation}
 
\textbf{Assumptions.} Despite the simple definition of the causal effect, the fundamental problem in causal inference, i.e., the \textit{missing data problem}, appear to be a major obstacle to the estimation of the causal effect. Therefore, it is critical to make additional assumptions.

\begin{assumption}[SUTVA]
\label{assumpt:sutva}
The potential outcomes for any unit do not vary with the treatments assigned to other units. For each unit, there are no different forms or versions of each treatment level, which lead to different potential outcomes.
\end{assumption}

The stable unit treatment value assumption, or SUTVA~\cite{imbens2015causal} is the most fundamental assumption in causal inference, incorporating both the \textit{No Interference} idea that treatments applied to one unit do not affect the outcome for another unit and the \textit{No Hidden Variations of Treatments} concept that for each unit there is only a single version of each treatment level. The second assumption, \textit{ignorability} or \textit{unconfoundedness}~\cite{rubin1990formal}, states that treatment assignment is free from dependence on the potential outcomes.

\begin{assumption}[Unconfoundedness / Ignorability]
\label{assumpt:uncon}
Treatment assignment $W$ is independent to the potential outcomes, i.e., $T \perp Y (T = 0),Y (T = 1)|X$, also written as $\textrm{Pr}(T=1|X, Y (T = 0),Y (T = 1)) = \textrm{Pr}(T=1|X)$, where $X$ denotes the background variables.
\end{assumption}

In other words, within subpopulations defined by the values of observed background variables, or covariates, the treatment assignment is random. The ignorability assumption rules out unmeasured confounders, which causally influences both the treatment $T$ and the outcome $Y(T)$. $\textrm{Pr}(T=1|X)$ is called the \textit{propensity score}~\cite{rosenbaum1983central}. The last assumption is positivity, or overlap:

\begin{assumption}[Positivity]
\label{assumpt:pos}
$0 < \textrm{Pr}(T=t|X=x) < 1, \forall t, x.$
\end{assumption}

In large data samples, positivity requires that there are both treated and control units for all values of the covariates. In contrast to the untestable ignorability assumption~\cite{ imbens2015causal}, positivity can be tested from observed data. The combination of unconfoundedness and positivity is referred to as “\textit{strong ignorability}~\cite{rosenbaum1983central}.”

\subsubsection{Structural Causal Models Framework}
\label{sec:scm}
Structural causal models (SCM)~\cite{pearl1995causal, pearl1988probabilistic, pearl2009causality} serve as a comprehensive causality framework, which unifies graphical models, nonparametric structural equations, and counterfactual and interventional logic. The most significant advantage of SCM is its intuitive structure of real-world causal dependencies based on graphical models as well as the wise and friendly symbiosis between counterfactual and graphical methods.

\textbf{Causal Graph.} A causal graph, or a causal diagram, is usually a Bayesian network, which describes the causal relations between variables by a Directed Acyclic Graph (DAG), where the nodes represent the variables and the edges record the causal relations. Causal graphs play an essential role in the SCM framework, for they provide a vivid representation of sets of variables that are relevant to each other in any given state of knowledge, and serves as a carrier of \textit{conditional independence} relationships along the order of construction, through which we can confirm whether it satisfies the criteria such that certain causal inference methods can be applied~\cite{pearl2009causality}. 

\begin{figure}[t!]
    \centering
    \vspace{-3mm}
    \subfloat[Chain]{
    	\centering
    	\includegraphics[width=0.2\textwidth, trim=-10 0 -10 0, clip]{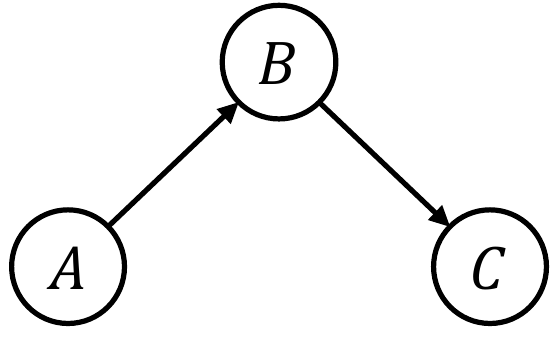}
    }
    \centering
    \subfloat[Fork]{
    	\centering
    	\includegraphics[width=0.2\textwidth, trim=-10 0 -10 0, clip]{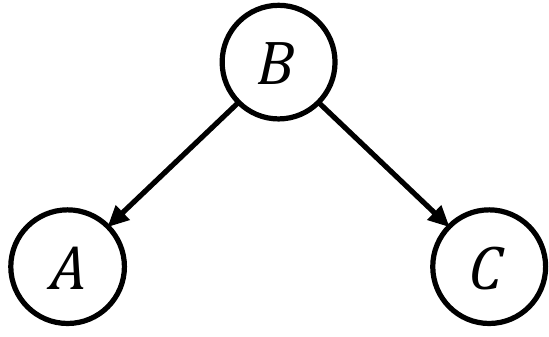}
    }
    \centering
        \subfloat[Collider]{
    	\centering
    	\includegraphics[width=0.2\textwidth, trim=-10 0 -10 0, clip]{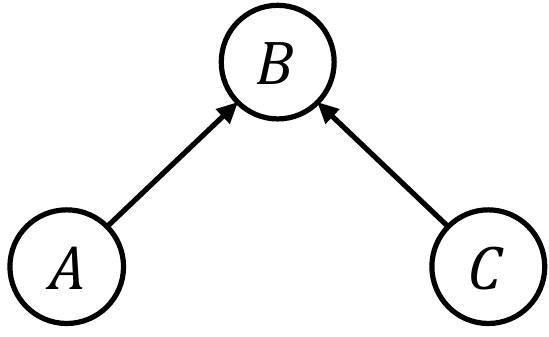}
    }
    \centering    
    \vspace{-3mm}
    \caption{Graphical models of three typical types of causal structures.}
    \vspace{-6mm}
    \label{fig:junction}
\end{figure}

\textbf{\textit{d}-Separation.} We first review the concept of \textit{dependency-separation} (\textit{d}-Separation) as the knowledge base for conditional independence. There are three typical causal graphs of three disjoint sets of variables, shown in Fig.~\ref{fig:junction}, with the help of which we can characterize any pattern of arrows in the network. In the \textit{chain} (Fig.~\ref{fig:junction}(a)), $B$ is the $mediator$ that transmits the effect of $A$ to $C$. In the \textit{fork} (Fig.~\ref{fig:junction}(b)), $B$ is often called a common cause or \textit{confounder} of $A$ and $C$. A confounder will make $A$ and $C$ statistically correlated even though there is no direct causal link between them, which may give rise to a so-called spurious correlation in the application. In the \textit{collider} (Fig.~\ref{fig:junction}(c)), though $A$ and $C$ are independent to begin with, conditioning on (i.e., knowing the value of) $B$ will make them dependent. A good example is three features of Hollywood actors: Talent $\rightarrow$ Celebrity $\leftarrow$ Beauty ~\cite{elwert2014endogenous}. Although beauty and talent are completely unrelated to one another in the general population, an unanticipated negative correlation is found between talent and beauty if we only focus on famous actors: a celebrity is unattractive increases our belief that he or she is talented~\cite{pearl2018book}. 
This negative correlation is sometimes called collider bias or the “explain-away” effect. 
In recommendation systems, a similar example can be found in analyzing popularity bias from the user's perspective. The factors influencing user interaction can be summarized as: Conformity $\rightarrow$ Interaction $\leftarrow$ Inherent Interest~\cite{zheng2021disentangling}. When a user interacts with a popular item, it does not necessarily indicate his or her true preference for it; such interaction may be driven by the desire to conform to prevailing trends. Conversely, if a user engages with an unpopular item — where the influence of conformity is significantly reduced — it is more probable that the item is in close alignment with the user's inherent interests.

A \textit{path} means a sequence of consecutive edges (of any directionality) in the graph, and we regard stopping the flow of dependency between the variables that are connected by such paths as \textit{blocked}. In the chain and fork, the path between $A$ and $C$ will be blocked by conditioning on $B$, while in the collider, any conditioning on $B$ will introduce a correlation between them. The formal definition of \textit{d}-separation or blocking is defined as follows.

\begin{definition}[\textit{d}-Separation]
\label{def:d_s}
A path is said to be d-separated (or blocked) by conditioning on a set of nodes $\mathcal{Z}$ if and only if one of the two conditions is satisfied:
\begin{enumerate}
\item The path contains a chain $A \rightarrow B \rightarrow C$ or a fork $A \leftarrow B \rightarrow C$ such that the middle node $B$ is in $\mathcal{Z}$;
\item The path contains a collider such that the middle node $B$ is not in $\mathcal{Z}$ and such that no descendant of $B$ is in $\mathcal{Z}$.
\end{enumerate}
\end{definition}

\begin{figure}[t!]
    \centering
    \vspace{-3mm}
    \subfloat[]{
    	\centering
    	\includegraphics[width=0.20\textwidth, trim=-4 0 -4 0, clip]{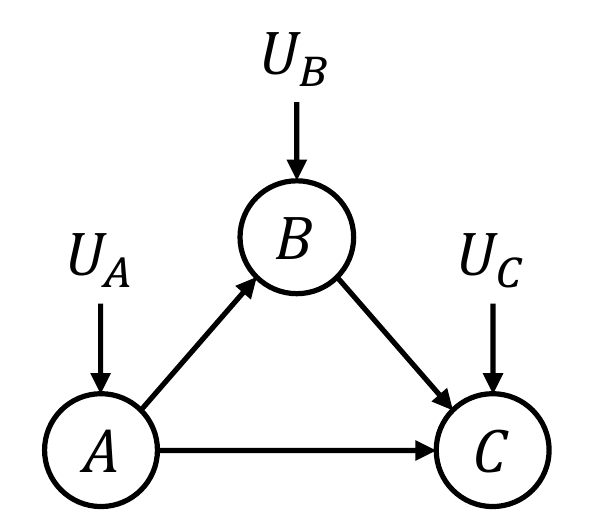}
    }
    \centering
    \subfloat[]{
    	\centering
    	\includegraphics[width=0.20\textwidth, trim=-4 0 -4 0, clip]{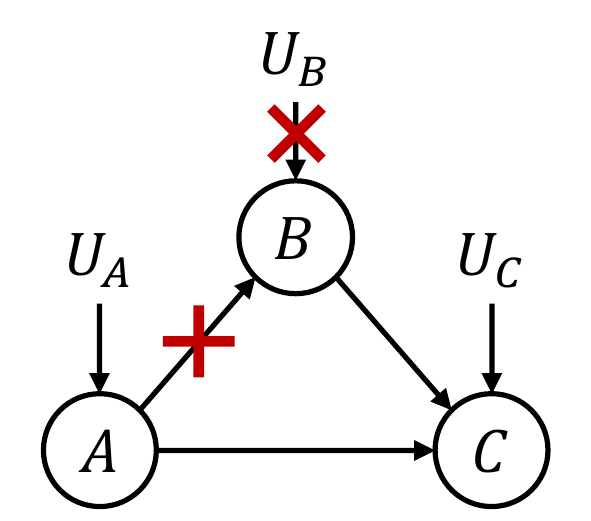}
    }
    \centering
    \vspace{-3mm}
    \caption{Examples of the structural equation and intervention.}
    \vspace{-7mm}
    \label{fig:inter}
\end{figure}

\textbf{Structural Equations.} Beside causal graph, structural equation is another representation of causal information, where the former is an abstraction of the latter. In its general form, a structural equation of a variable $Y$ is defined as:

\begin{equation}
\label{eq:sem}
Y=f_Y(Pa, U),
\end{equation}
where $Pa$ (connoting parents) stands for the set of variables that directly determine the value of $Y$ and where $U$ represents exogenous variables or errors (or “disturbances”) due to omitted factors. For example, the causal graph in Fig.~\ref{fig:inter}(a) is associated with the structural model as follows. In this section, uppercase letters are used to represent variables, while their corresponding lowercase counterparts denote the values of these variables. Thus, we have:
\begin{align}
	\label{eq:structural_example}
	\begin{cases}
	a&=\ f_A(u_A), \\
	b&=\ f_B(a, u_B), \\
	c&=\ f_C(a, b, u_C),
	\end{cases}
\end{align}
where $U_A$, $U_B$ and $U_C$ represent exogenous variables. A set of equations in the form of Equation~\ref{eq:sem} is called a \textit{structural model}; if each variable has a distinct equation in which it appears on the left-hand side, then the model is called a \textit{structural causal model}.

\textbf{Intervention.} The \textit{do-calculus} allows researchers to complete \textit{intervention}, interpreted as controlling the value of a variable, by purely mathematical means instead of by carrying out a physical experiment, which is one of the outstanding contributions of Pearl’s SCM framework. The \textit{do}-calculus involves the do-operation, like $do(T=t)$, which denotes the intervention of setting the variable $T$ to $t$, realizing by blocking the effect of $T$’s parents on $T$ and set the value of $T$ as $t$. For example, if we $do(B=b_0)$ on the model in Fig.~\ref{fig:inter}(a), Equations~\ref{eq:structural_example} will be modified as:
\begin{align}
	\label{eq:intervention}
	\begin{cases}
	a&=\ f_A(u_A), \\
	b&=\ b_0, \\
	c&=\ f_C(a, b_0, u_C), 
	\end{cases}
\end{align}
the graphical description of which is shown in Fig.~\ref{fig:inter}(b).

It is crucial to note that $\textrm{Pr}(Y=y|do(T=t))$ and $\textrm{Pr}(Y=y|T=t)$ are not the same. For example. the fork structure (Fig.\ref{fig:junction}(b))might represent the causal mechanism that connects the number of sales at a local ice cream shop on that day ($A$), a day's temperature in a city ($B$), and the number of violent crimes in the city on that day ($C$)~\cite{glymour2016causal}. Because both ice cream sales and violent crime are more common in hot weather, a positive correlation might be found when estimate $P(C=c|A=a)$. However, as illustrated in manipulated graphical model of Fig.~\ref{fig:inter}, crime rates $C$ are independent of ice cream sales $B$, which results in a different $\textrm{Pr}(C=c|do(A=a))$ from $\textrm{Pr}(C=c|A=a)$.

Although causal graph manipulation is the most fundamentalist approach to calculating $\textrm{Pr}(Y=y|do(T=t))$, it can be challenging and even impossible in reality. Fortunately, we can estimate $\textrm{Pr}(Y=y|do(T=t))$ from observed data with the following causal effect rule:
\begin{definition}[The Causal Effect Rule]
\label{def:ce_rule}
Given a graph $G$ in which a set of variables $PA$ are designated as the parents of $T$, the causal effect of $T$ on $Y$ is given by
\begin{equation}
\label{eq:ce_rule}
\textrm{Pr}(Y=y \mid do(T=t)) 
= \sum_{x} \textrm{Pr}(Y=y \mid T=t, PA=x) \textrm{Pr}(PA=x) 
= \sum_{x} \frac{\textrm{Pr}(T=t, Y=y, PA=t)}{\textrm{Pr}(T=t \mid PA=x)}, 
\end{equation}
where $x$ ranges over all the combinations of values that the variables in $PA$ can take.
\end{definition}
The most important benefit brought by the rule is that it enables us to finish the  \textit{do}-calculus purely on passive observational data~\cite{xu2021causal}. The factor $\textrm{Pr}(T=t \mid PA=x)$ is the propensity score, and Equation \ref{eq:ce_rule} is named \textit{inverse propensity score} (Section \ref{sec:ips}) in PO framework, which partly reflects the unity of the two frameworks.


\textbf{Counterfactuals.} Counterfactuals are employed to emphasize our wish to compare two outcomes under the exact same conditions, differing only in one aspect: the \textit{antecedent}, or \textit{hypothetical condition}~\cite{glymour2016causal}. For example, in the counterfactual question “What would be the user's interaction if the recommended items had been different?” mentioned above in the Section \ref{sec:intro}, we would like to compare the user’s interaction under the same conditions except for the recommended item. Counterfactuals, situations which are non-existent in reality, cannot be inferred by do-calculus. Fortunately, Pearl~\cite{pearl2009causality} proposed a new set of notations: $ \textrm{Pr}(Y(T=1)|T = 0, Y = Y(T=0))$ indicates the probability of the outcome $Y(T=1)$ would be if the observed treatment value is $T=0$, given the fact that we observe $Y=Y(T=0)$ in the data.

\subsubsection{Comparison between the two frameworks}
As mentioned above, the two frameworks are equivalent logically: an assumption or a theorem can be translated to its counterpart in the other,  and a problem solved in one framework would yield the same solution in another~\cite{pearl2009causality, glymour2016causal}. For example, $\textrm{Pr}(Y = y | do(T = t))$ in the SCM is equivalent to $\textrm{Pr}(Y(T=t)=y)$ in the PO, which the regular assessment in a controlled experiment, in which the distribution of $Y$ is estimated for each level $w$ of a random variable $T$. Causal effects that are measured between the results of the counterfactual world and the real world can be estimated conveniently in both frameworks. However, there are several important differences between PO and SCM. The most significant difference is that PO does not assume the causal relations between concerned variables, while SCM makes assumptions of causal mechanisms among a set of variables or searches for ones based on some assumptions. In other words, any given PO model corresponds to multiple causal graphs in SCM. For PO, it can be a strength for PO that causal effects can be reasoned without knowing the causal model, and be a weakness either. According to the unconfoundednes assumption, all confounders should be observed to infer a correct treatment effect since the mechanism is unknown, almost impossible in practice~\cite{aliprantis2015distinction}. In contrast, in SCM, causal diagrams allow us to work with causal effects by interventions on the fewest number of variables or the observed variables as much as possible.

\subsection{Recommender Systems}

Recommender systems predict users’ preferences and proactively recommend items users might like~\cite{ricci2015recommender, zhang2019deep} to alleviate information overload. 

\subsubsection{Recommendation Techniques}

RSs are usually classified into the following three categories~\cite{adomavicius2005toward, zhang2019deep}: content-based, collaborative filtering (CF), and hybrid. Content-based recommendation learns to recommend primarily based on comparisons across items’ and users’ auxiliary information~\cite{zhang2019deep}, such as items’ human-set tags, images, texts, and users’ sex. Collaborative filtering recommender systems recommend items according to user/item historical interactions, i.e., explicit (e.g., user’s previous ratings) or implicit feedback (e.g., click behavior)~\cite{zhang2019deep}. Hybrid approaches are those that combine collaborative filtering and content-based methods.

If we review the model structure, recommender systems can generally be divided into shallow models and neural network models. Shallow models involve methods that directly calculate the similarity of interactions and CF methods with matrix factorization (MF)~\cite{koren2009matrix} or factorization machine (FM) ~\cite{rendle2010factorization}, but suffer from insufficient learning of users’ complicated interest. Neural network models are proposed to solve this issue, with the advantage of high-order feature interactions~\cite{guo2017deepfm}. For example, Wide \& Deep~\cite{cheng2016wide} jointly trains linear models and deep neural networks to combine the benefits of memorization and generalization. Deep factorization machine (DeepFM)~\cite{guo2017deepfm} combines traditional  factorization machine (FM) with multi-layer perceptrons (MLP) in parallel. Graph neural networks (GNN)-based methods adopt embedding propagation to iteratively aggregate neighborhood embedding, thereby more effectively exploring structural information~\cite{gao2022graph}.

\subsubsection{Notation}

Considering a general recommender system, we assume $\mathcal{O}$ and $\mathcal{O}^{-}$ denote the observed dataset and unobserved dataset. Each observed sample includes the treatment $T$, background features $X$, and an interaction label $Y$. Background features $X$, aka., covariates are usually formulated as a high dimensional sparse vector containing information such as user ID, item ID, user profile, item category, etc. The interaction label $Y$, or the outcome, can be explicit feedback (e.g., rating) or implicit feedback (such as click and watch behavior). 

In normal circumstances, researchers prefer choosing whether to recommend as the treatment. Therefore, the observed dataset can be denoted as ${\mathcal{O}} = \{(T=1,X,Y)\}|_{1}^{|\mathcal{O}|} \in \mathcal{T} \times \mathcal{X} \times \mathcal{Y}$, where $\mathcal{T}$ means the treatment space, $\mathcal{X}$ is the feature spaces, and $\mathcal{Y}$ is the label space. In general, the observed dataset is obtained with the deployed recommender policy $\pi$; thus ${\mathcal{O}}$ will be specifically expressed as ${\mathcal{O}_\pi}$ if we are concerned about the policy. Note that settings of $T, X, Y$ vary slightly according to specific work. It would be better to understand with reference to the context.

\subsection{Existing Categorizations of Causal Recommendation}
\label{sec:exist_cate}
There are several categorization criteria for causal recommender systems. For example, similar to ~\cite{yao2021survey}, Yao et.al.~\cite{wu2022opportunity} divides biases in RS into three categories from the perspective of violating what causal assumptions are adopted in the standard PO framework. 1) Position bias and conformity bias can be seen as violations of the SUTVA assumption if recommender systems do not pay enough attention to the positions of items and users’ social networks. 2) Unconfoundedness and positivity are crucial assumptions in the recoverability of the target estimated. However, the former can be violated by popularity bias, and the latter can be violated by exposure bias, both of which result in the problem of missing not at random (MNAR). 3) The final bias violates some model-specific assumptions.

According to the survey~\cite{gao2022causal}, existing work of causal recommendation can be categorized into three groups: for addressing data bias, for addressing data missing and noise, and for beyond-accuracy objectives. 1) Causal debiasing work can be further divided into several subcategories based on the specific bias, such as popularity bias, clickbait bias, and exposure bias. 2) The problem of data missing refers to the usually-discussed data sparsity issue in RS, and data noise stems from unreliable implicit signals and delayed feedback. In order to alleviate these issues, researchers use the counterfactual technique to augment insufficient data and adjust sample weights. Besides, some causal recommender systems are designed for beyond-accuracy objectives like explainability, diversity and fairness. 

Zhu et al.~\cite{zhu2023causal} summarizes different causal inference techniques with an emphasis on debiasing, explainability promotion, and generalization improvement. Xu et al.~\cite{xu2023causal} introduces existing causal methods on explainable recommendation, fairness in recommendation, uplift-based recommendation, robust recommendation, unbiased recommendation, respective.


The four studies mentioned above are pioneering efforts in this field, and each has a distinct focus on the causal frameworks it discussed. However, this paper will systematically classify causal inference for RS from a new perspective of the employed causal theories. We regard that, while it is undeniably convenient for researchers, especially those embedded in the RS industry, to quickly reference existing causal methods based on the application issues they address, these categorizations result in a fragmented and non-systematic representation of causal theories, since a single causal theory could potentially be applied to resolve a variety of recommendation issues.

Consequently, this paper systematically classifies causal inference for RS from a new perspective of the employed causal approach. This taxonomy enables readers to grasp the progressive integration and iterative development of causal methods within RS, fostering an understanding of their advantages over previous techniques, as well as their inherent limitations. Such a comprehensive overview is instrumental for continuous research and paves the way for significant breakthroughs in the in-depth integration of causal inference and recommender systems, ensuring a more robust and holistic development in the field.
\section{PO-based Methods}
\label{sec:po_based}

Many causal recommendation approaches, especially in early research, have focused on applying the potential outcome (PO) framework proposed by Donald B. Rubin ~\cite{splawa1990application, rubin1974estimating, imbens2015causal}.
These approaches primarily integrate PO-based causal inference into the optimization functions in traditional deep-learning-based methods or the reward functions in reinforcement-learning-based methods. 

Fig.~\ref{fig:classification} illustrates the strategies and objectives concerning the PO framework in the context of RS, categorizing the strategies into two main types: propensity score and causal effect. The former generally leverages estimated propensity scores from causal inference methods to adjust importance weights, while the latter concentrates on the difference between potential outcomes under treatment and control (see Definition \ref{eq:ITE}). Despite their different focuses, they are not entirely mutually exclusive. On one hand, propensity scores can be utilized to adjust the weights of samples or the weights of outcomes within causal effects. On the other hand, causal effects can be estimated in a couple of ways. One approach involves directly modeling outcomes, exemplified by fitting two separate models~\cite{radcliffe2007using, bonner2018causal} to estimate $\mathbb{E}[Y(T=1|X=x)]$ and $\mathbb{E}[Y(T=0|X=x)]$ in the CATE (refer to Equation~\ref{eq:CATE}). Alternatively, propensity score-based methods like Inverse Propensity Scoring (IPS) or Doubly Robust (DR) can be applied to weigh the potential outcome predictions.

It is essential to clarify that in this paper, models that estimate causal effects without explicitly utilizing causal structure information are classified as PO-based, whereas those explicitly incorporating causal structure information are categorized as SCM-based. 

\subsection{Propensity Score Strategy}

Let's consider the process by which the recommendation system works, where given background variables $x \sim \textrm{Pr}(x)$, also referred to as pre-treatment variables or covariates~\cite{imbens2015causal}, (e.g., user and item features, time of the day, etc.), a recommender policy $\pi$ plays a role as a decision-making system, which makes a decision of whether to take an active treatment $t \sim \pi(t \mid x) $ (e.g., recommend an item), and the potential outcome $y \sim \textrm{Pr}(y \mid x, t) $, i.e., “reward” in the reinforcement learning context (e.g., click indicator), will be observed ~\cite{saito2022off}. For example, in online markets, information like user profile, historical consumptions, and products in the cart will be treated as context variables $x$, according to which the policy $\pi$ will produce a list of recommended items (i.e., treatment $t$), and the logged reward $y$ can be the click signal, conversions, or revenue, etc. The effectiveness of the policy $\pi$ can be evaluated through its running expected reward, formulated as:
\begin{equation}
R(\pi) :=\iiint y \textrm{Pr}(y \mid x, t) \pi(t \mid x) \textrm{Pr}(x) d x d t d y
=\mathbb{E}_{\textrm{Pr}(x) \pi(t \mid x) \textrm{Pr}(y \mid x, t)}[y].
\end{equation}

To learn the optimal policy 

\begin{equation}
\pi \in \underset{\pi \in \mathcal{\Pi}}{\arg \max } V(\pi),
\end{equation} 
where $\Pi$ means the policy class, an online A/B test will be the best choice~\cite{gomez2015netflix, kohavi2013online}, but   suffers from high expense. A substitute and common practice is offline evaluation,  by calculating an estimator $\hat{R}$ for the reward of a target policy $\pi$ using logged data $\mathcal{O}_{\pi_0}$ collected by a logging policy $\pi_0$ (which is different from $\pi$)~\cite{saito2022off}. However, like many other empirical sciences, offline evaluation is challenged with the problem of \emph{missing not at random} (MNAR). 

To address this issue, early approaches tend to predict the missing data directly~\cite{steck2010training} but have accentuated the problem of high bias~\cite{wang2019doubly, saito2021counterfactual}. Recently, many researchers have resorted to the \emph{propensity score} $e(X)$ in causality to recover the data distribution. For example, ExpoMF~\cite{liang2016modeling} first predicts the exposure matrix and then uses the exposures (i.e., propensity scores) to guide the model of the interaction matrix, which is inspired by the separation between propensity scores and potential outcomes in the PO framework. Similarly, Wang et al.~\cite{wang2018collaborative} propose SERec to integrate social exposure into collaborative filtering. A refreshing work is that Wang et al.~\cite{wang2020causal} aim to overcome the confounder issue with propensity score. They regard correlations among the interacted items as bringing indirect evidence for confounders and propose the deconfounded recommenders. They first build an exposure model to estimate the propensity score, and then use this exposure model to estimate a substitute for the unobserved confounders, conditional on which the final outcome model (specifically in ~\cite{wang2020causal}, a rating model based on matrix factorization) is trained. In addition, inspired by ~\cite{joachims2017unbiased, fang2019intervention}, Chen et al.~\cite{chen2021adapting} propose IOBM (Interactional Observation-Based Model)to estimate propensity score in interaction settings, which learns low-dimensional embeddings as a substitute for unobservable confounders. Specifically, it learns individual embeddings to capture the potential outcome information from specific exposure events. Based on individual embeddings, the interactional embeddings, which uncovers the hidden relationship among single exposure events and utilizes query context information to apply attention, are learned through the bidirectional LSTM model. Recently, the incorporation of Contrastive Learning (CL)~\cite{yu2023self, zhou2021contrastive} with propensity scores has offered new avenues to address noisy data in recommendation systems. A prominent example is the CCL (Contrastive Causal Learning) framework~\cite{zhou2023contrastive}, which innovatively employs propensity score-based sampling to generate informative positive pairs for contrastive learning tasks.

\begin{figure*}[t!]
\centering
\vspace{-1mm}
\includegraphics[height=0.55\textheight,trim=10 0 10 0 ,clip]{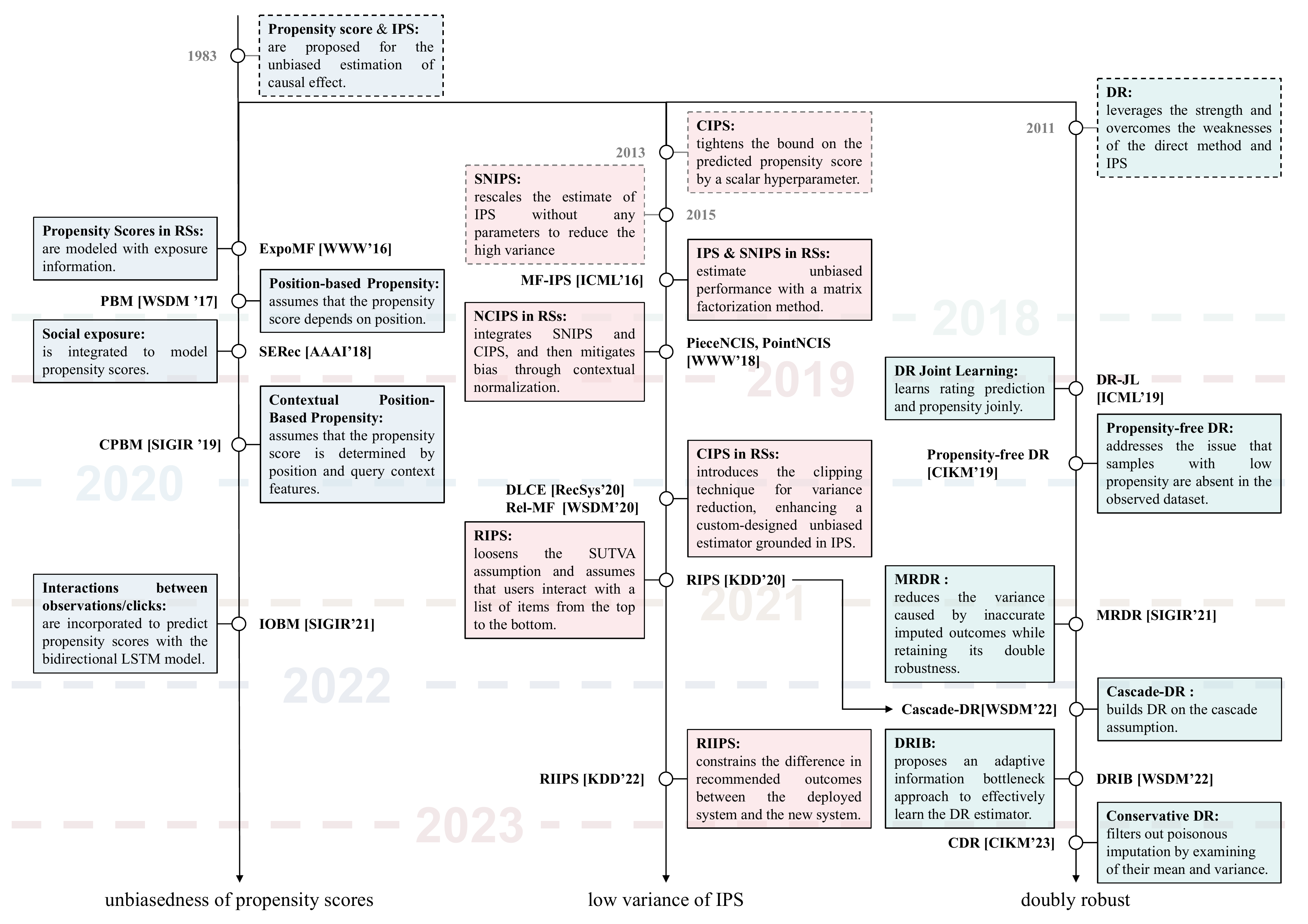} 
\vspace{-3mm}
\caption{Evolutionary Timeline of Propensity Score Strategies in Recommendations.}
\vspace{-6mm}
\label{fig:propensity}
\end{figure*}

Propensity-based methods can be further divided into approaches based on inverse propensity score (IPS) and approaches based on doubly robust (DR) (Fig.\ref{fig:propensity}). One of the greatest strengths of applying propensity-based methods in RS is that most of them are unbiased and model-agnostic, simply deployed on the objective function for policy evaluation directly or for policy learning indirectly.

\subsubsection{Missing Not At Random}
\label{sec:MNAR}

In this part, we will introduce the phenomena and factors of missing not at random, to provide explanations and conclusions of challenges in recommender systems in a causal language to understand existing work better.

Recommendation algorithms often obey the missing at random (MAR)~\cite{rubin1976inference} assumption but may lead to biased prediction and suboptimal policy~\cite{little2019statistical, marlin2009collaborative}. The MAR condition essentially states that the probability that a potential outcome is missing does not depend on the value of that potential outcome and can be easily violated in recommender systems~\cite{marlin2009collaborative}. For example, on movie rating websites, movies with high ratings are less likely to be missing compared to movies with low ratings~\cite{pradel2012ranking}. The issue of missing not at random (MNAR) has been demonstrated by Marlin and Zemel ~\cite{marlin2009collaborative} and it is a phenomena stemming from \emph{selection bias} and \emph{confounding bias}~\cite{correa2019identification, wu2022opportunity}. 

\begin{figure}[t!]
    \centering
    \vspace{-3mm}
    \subfloat[User self-selection bias]{
    	\centering
    	\includegraphics[height=0.09\textheight, trim=-10 0 -10 0, clip]{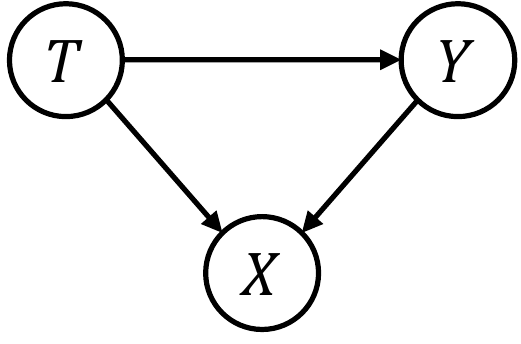}
    }
    \centering
    \subfloat[Confounding bias]{
    	\centering
    	\includegraphics[height=0.09\textheight, trim=-10 0 -10 0, clip]{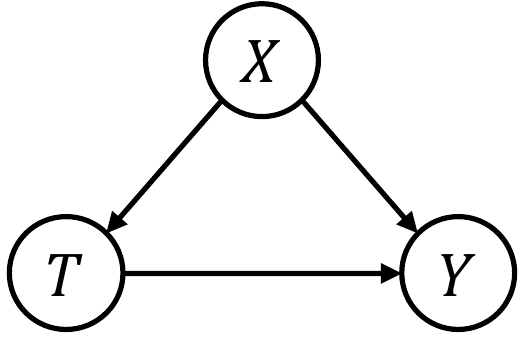}
    }
    \centering
    \vspace{-3mm}
    \caption{Causal explanation of user self-selection bias and confounding bias.}
    \vspace{-6mm}
    \label{fig:selection_confounding}
\end{figure}

Selection bias, or sampling bias, is usually discussed in the prediction task and can be further classified into model selection bias and user self-selection bias~\cite{wu2022opportunity}. For example, the case that the platform may systematically recommend pop music to younger users who may be more active on the service regardless of genre preferences~\cite{mcinerney2020counterfactual} will be regarded as model selection bias~\cite{yuan2019improving, mcinerney2020counterfactual} and can be eliminated by random recommendation. User self-selection bias~\cite{bareinboim2012controlling, elwert2014endogenous}, on the contrary, can not be removed by randomization of recommendation~\cite{correa2019identification}. It is caused by preferential exclusion of samples from the data~\cite{bareinboim2012controlling}. A typical example is a song recommender system, in which users usually rate songs they like or dislike and seldom rate what they feel neutral about~\cite{saito2020asymmetric}. Some of the most frequently discussed biases like popularity bias~\cite{zhang2021causal, wei2021model} and exposure bias~\cite{liang2016modeling, wang2018collaborative} will lead to model selection bias, while conformity bias~\cite{zhang2021causal, zheng2021disentangling} and clickbait bias~\cite{wang2021clicks} fall under user self-selection bias as a result of user preference.

Confounding bias~\cite{hernan2002causal, pearl2009causality} arises from the confounder described in Section \ref{sec:scm}, which affects both the treatment and the outcome, illuminated in Fig.\ref{fig:selection_confounding}(b). Alternatively, it can be identified if the probabilistic distribution representing the statistical association is not always equivalent to the interventional distribution, i.e., $\textrm{Pr}(y \mid t) \neq \textrm{Pr}(y \mid do(t))$~\cite{guo2020survey}. A notable example of confounding bias is that a system trained with historical user interactions may over recommend items that the user used to like, and the user’s decision (i.e., outcome) is also affected by historical interactions~\cite{wang2021deconfounded}.

Both biases can lead to invalid estimates of causality from the data, and they are not mutually exclusive because selection bias does not explicitly involve causality. Many model selection biases, including popularity bias and exposure bias, are also confounding biases. As for user self-selection bias, the model in Fig. \ref{fig:selection_confounding} (a) gives an illustration of its causal nature in which $S$ is a variable affected by both $T$ (treatment) and $Y$ (outcome), indicating entry into the data pool~\cite{bareinboim2012controlling}. Therefore, confounding bias is significantly different from user self-selection bias from the causal perspective. The former originates from common causes, whereas the latter originates from common outcomes~\cite{elwert2014endogenous}. The former stems from the systematic bias introduced during the treatment assignment, while the latter comes from the systematic bias during the collection of units into the sample~\cite{correa2019identification}.

\subsubsection{Inverse Propensity Score}
\label{sec:ips}
Inverse Propensity Score (IPS)~\cite{horvitz1952generalization, rosenbaum1987model, rosenbaum1983central,little2019statistical}, also named as inverse propensity weighting (IPW), or inverse propensity of treatment weighting (IPTW), is one of the favorite counterfactual techniques and has inspired a lot of causal inference methods in RS, especially for unbiased learning~\cite{joachims2017unbiased}. Propensity score is the probability of receiving the treatment given covariates $X$, formulated as:
\begin{equation}
e_{\pi}(X) = \textrm{Pr}_{\pi}(T=1 \mid X).
\end{equation}
IPS assigns a weight $w$ to each sample:
\begin{equation}
w=\frac{t}{e(x)}+\frac{1-t}{1-e(x)},
\end{equation}
which indicates the inverse probability of receiving the \emph{observed} treatment and control. The unbiasedness of IPS can be proven~\cite{rosenbaum1987model}. More specifically, for the reward estimation of recommendation policy, IPS adjusts the distribution of background features in the logged dataset to be consistent with that during $\pi$ tests online, formulated as:
\begin{equation}
\hat{R}_{\mathrm{IPS}}\left(\pi; \mathcal{O}_{\pi_0}\right):=\frac{1}{\mathcal{O}_{\pi_0}} \sum_{k=1}^{|\mathcal{O}_{\pi_0}|} \frac{e_{\pi}(X)}{e_{\pi_0}(X)} \cdot y_k = \frac{1}{\mathcal{O}_{\pi_0}} \sum_{k=1}^{|\mathcal{O}_{\pi_0}|} \frac{\textrm{Pr}_{\pi}(T=1 \mid X)}{\textrm{Pr}_{\pi_0}(T=1 \mid X)} \cdot y_k,
\end{equation}
where we assume that only positive feedback is taken into account, and $w= \frac{e_{\pi}(X)}{e_{\pi_0}(X)} $ is the ratio of the evaluation and logged policies. Note that in most applications in RS, IPS is model-agnostic, applied to the training objective function for policy evaluation directly or for policy learning indirectly.

\begin{table*}[pt]
  \small
  \centering
  \vspace{-1mm}
  \caption{\normalsize Summary of propensity score strategies for recommendation.}
  \vspace{-2mm}
  \setlength{\tabcolsep}{1.5mm}{
    \begin{tabular}{c|c|c|c|c|c}
    \toprule
    \textbf{Category} & \textbf{Model} & \textbf{Causal method} & \textbf{Backbone model} & \textbf{Issue of concern} & \textbf{Year} \\
    \midrule
    \multicolumn{1}{c|}{\multirow{6}[2]{*}{\makecell{Approach \\ Inspired by \\Propensity\\ Score}}} & ExpoMF~\cite{liang2016modeling} & Propensity score & MF  & Exposure bias & 2016 \\
          & SERec~\cite{wang2018collaborative} & Propensity score & MF  & Social recommendation & 2018 \\
          & Dcf~\cite{wang2020causal} & Propensity score & MF  & Unobserved confounding bias  & 2020 \\
          & CNFI~\cite{zhang2021causalneural} & Propensity score & MF    & Implicit feedback & 2021 \\
          & IOBM~\cite{chen2021adapting} & Propensity score & Bi-LSTM~\cite{graves2013speech}  & Interactional observation bias & 2021 \\
          & CCL~\cite{zhou2023contrastive} & Propensity score & (custom-designed) & Unobserved confounding bias  & 2023 \\
    \midrule
    \multirow{19}[2]{*}{\makecell{Approach\\ with Inverse\\ Propensity \\Score (IPS)}} & MF-IPS~\cite{schnabel2016recommendations} & IPS, SNIPS & MF  & Selection bias & 2016 \\
          & PBM~\cite{joachims2017unbiased} & IPS   & SVM-Rank~\cite{joachims2002optimizing, joachims2006training} & Position bias & 2017 \\
          & PieceNCIS, PointNCIS~\cite{gilotte2018offline} & CIPS, SNIPS & - & Offline A/B testing & 2018 \\
          & ~\cite{mehrotra2018towards} & IPS   & (reinforcement learning) & Fairness & 2018 \\
          & Multi-IPW~\cite{zhang2020large} & IPS   & Multi-task MLP & Selection bias & 2019 \\
          & CPBM~\cite{fang2019intervention} & IPS   & SVM-Rank & Selection bias & 2019 \\
          & ULRMF,ULBPR~\cite{sato2019uplift} & IPS, SNIPS, ATE   & MF  & Uplift & 2019 \\
          & DLCE~\cite{sato2020unbiased} & CIPS  & MF  & Unobserved confounding bias  & 2020 \\
          & Rel-MF~\cite{saito2020unbiased} & CIPS  & MF  & Unobserved confounding bias  & 2020 \\
          & ~\cite{christakopoulou2020deconfounding} & IPS   & Multi-task DNN & Observed confounding bias & 2020 \\
          & RIPS~\cite{mcinerney2020counterfactual} & RIPS  & (model-agnostic) & Slate recommendation & 2020 \\
          & ACL-~\cite{xu2020adversarial} & IPS   & \makecell{(adversarial learning)} & Identifiability & 2020 \\
          & UR-IPW~\cite{zhang2021user} & SNIPS & Multi-task MLP & \makecell{Post-click revisit effect\\ \&selection bias} & 2021 \\
          & ~\cite{li2021debiasing} & IPS   & (model-agnostic) & Domain bias & 2021 \\
          & CBDF~\cite{zhang2021counterfactual} & IPS   & (reinforcement learning) & Delayed feedback & 2021 \\
          & RD\&BRD~\cite{ding2022addressing} & \makecell{IPS/DR/\\ AutoDebias~\cite{chen2021autodebias}} & MF  & Unobserved confounding bias  & 2022 \\
          & CET~\cite{cai2022hard} & IPS   & BERT  & False negative  & 2022 \\
          & CAFL~\cite{krauth2022breaking} & IPS   & MF  & Feedback loop & 2022 \\
          & RIIPS~\cite{liu2022practical} & RIIPS  & Two-tower structure  & Selection bias & 2022 \\
          & DENC~\cite{li2022causal} & IPS   & (custom-designed) & Selection bias & 2023 \\
    \midrule
    \multirow{10}[2]{*}{\makecell{Approach \\with Doubly\\ Robust}} & Propensity-free DR~\cite{yuan2019improving} & DR    & FFM~\cite{yuan2019one} & Selection bias & 2019 \\
          & DR-JL~\cite{wang2019doubly} & DR & MF & Selection bias & 2019 \\
          & Multi-DR~\cite{zhang2020large} & DR    & Multi-task MLPDNN & Selection bias & 2020 \\
          & MRDR-DL~\cite{guo2021enhanced} & MRDR  & MF  & Selection bias & 2021 \\
          & Cascade-DR~\cite{kiyohara2022doubly} & Cascade-DR & MF  & High variance of RIPS~\cite{mcinerney2020counterfactual} & 2022 \\
          & ASPIRE~\cite{mondal2022aspire} & DR, ATE    & LightGBM~\cite{ke2017lightgbm} & Uplift & 2022 \\
          & DRIB~\cite{xiao2022towards} & DR    & MF  & Unobserved confounding bias  & 2022 \\
          & DR-BIAS, DR-MSE~\cite{dai2022generalized} & DR & FM & Selection bias & 2022 \\
          & CDR~\cite{song2023cdr} & DR & MF & Selection bias & 2023 \\
          & CF-MTL~\cite{li2023should} & CATE, IPS, DR & (custom-designed) & Personalized incentive policy & 2023 \\
    \bottomrule
    \end{tabular}}%
    \label{tab:propensity}%
    \vspace{-6mm}
\end{table*}%

Much IPS-based recommendation focuses on data debiasing in user interactions, mainly selection bias ~\cite{schnabel2016recommendations,saito2020unbiased,sato2020unbiased,zhang2021user,sato2021online, zhang2021causalneural, wu2021unbiased, li2022causal}. For example, ~\cite{schnabel2016recommendations} is a representative work adopting IPS to recommender system for the elimination of selection bias, in which the recommendation algorithm is based on matrix factorization and propensity scores are estimated via naive Bayes or logistic regression. Similarly, Saito et al.~\cite{saito2020unbiased} estimate the exposure propensity for each user-item pair and Sato et al.~\cite{sato2020unbiased} propose the DLCE (Debiased Learning for the Causal Effect) model with IPS-based estimators to evaluating unbiased ranking uplift. Unbiased IPS-based uplift is also concerned by ~\cite{sato2019uplift}. In addition, ~\cite{zhang2021user} proposes UR-IPW (User Retention Modeling with Inverse Propensity Weighting) to model revisit rate estimation accounting for the selection bias problem and ~\cite{li2021debiasing} adjusts domain weights based on IPS to reduce domain bias. Though IPS-based methods do not require an explicit analysis of the causal correlation between variables, some works~\cite{christakopoulou2020deconfounding, mcinerney2020counterfactual, ding2022addressing} still discuss causal graphs as an excellent guide to accurate model. For example, Ding et al.~\cite{ding2022addressing} leverage a causal graph to explain the risk of unmeasurable confounders on the accuracy of propensity estimation and propose RD (Robust Deconfounder) with the sensitivity analysis, obtaining the bound of propensity score to enhance the robustness of methods against unmeasured confounders. Li et al.~\cite{li2022causal} construct the DENC (De-bias Network Confounding in Recommendation). This causal graph-based recommendation framework disentangles three determinants for the outcomes, including inherent factors, social network-based confounder and exposure, and estimates each of them with a specific component, respectively. 
By the way, there are some works~\cite{christakopoulou2020deconfounding, cai2022hard, zhang2021user} integrate multi-task models with IPS to learn propensity scores and user interactions simultaneously.

In addition to debiasing, some IPS-based methods are dedicated to addressing other issues that abound in RS~\cite{mehrotra2018towards, zhang2021counterfactual, krauth2022breaking}. For example, Mehrotra et al.~\cite{mehrotra2018towards} proposes an unbiased estimator of user satisfaction based on IPS to jointly optimize for supplier fairness and consumer relevance. Besides, the CBDF (Counterfactual Bandit with Delayed Feedback) algorithm ~\cite{zhang2021counterfactual} re-weights the observed feedback with importance sampling, which is determined by a survival model to deal with delayed feedbacks. The CAFL (causal adjustment for feedback loops) ~\cite{krauth2022breaking} extends the IPS estimator to break feedback loops. 

Despite the unbiasedness strength of IPS, the inaccurate estimation of the unknown propensity $e(x)$ or sample weight, which results in high variance~\cite{gilotte2018offline}, becomes the biggest obstacle to achieving it. To alleviate this problem, modified versions of IPS have been proposed to control variance and applied to RS, including Self Normalized IPS ~\cite{schnabel2016recommendations, zhang2021user}, Clipped IPS~\cite{saito2020unbiased, sato2020unbiased}, Reward interaction IPS~\cite{mcinerney2020counterfactual}, and Regularized per-Item IPS~\cite{liu2022practical}. Self Normalized Inverse Propensity Scoring (SNIPS)~\cite{swaminathan2015self} rescales the estimate of the original IPS without any parameters to reduce the high variance, which is:
\begin{equation}
\hat{R}_{\mathrm{SNIPS}}\left(\pi; \mathcal{O}_{\pi_0}\right):=
\left(\sum_{k=1}^{|\mathcal{O}_{\pi_0}|} \frac{e_{\pi}(X)}{e_{\pi_0}(X)}\right)^{-1}
\sum_{k=1}^{|\mathcal{O}_{\pi_0}|} \frac{e_{\pi}(X)}{e_{\pi_0}(X)} \cdot y_k,
\end{equation}
and is introduced to RS by works like ~\cite{schnabel2016recommendations} and ~\cite{zhang2021user} to alleviate selection bias. Clipped IPS (CIPS)~\cite{bottou2013counterfactual, saito2020unbiased, sato2020unbiased}, or Capped IPS, tightens the bound of the sample weight by introducing a scalar hyperparameter $\lambda_{\mathrm{CIPS}}$, formulated as:
\begin{equation}
\hat{R}_{\mathrm{CIPS}}\left(\pi; \mathcal{O}_{\pi_0}\right):=
\frac{1}{\mathcal{O}_{\pi_0}}
\sum_{k=1}^{|\mathcal{O}_{\pi_0}|} \min \left\{ \frac{e_{\pi}(X)}{e_{\pi_0}(X)} , \lambda_{\mathrm{CIPS}} \right\}\cdot y_k,
\end{equation}
which has a lower variance but gives away its unbiasedness. Expanding upon the groundwork established by NCIPS~\cite{swaminathan2015self}, which amalgamated SNIPS and CIPS, the study by Gilotte et al.~\cite{gilotte2018offline} advances PieceNCIS and PointNCIS as enhancements that utilize contextual information to refine bias modeling. McInerney et al.~\cite{mcinerney2020counterfactual} loosen the SUTVA assumption and propose Reward interaction IPS (RIPS) for sequential recommendations, which assumes a causal model in which users interact with a list of items from the top to the bottom. RIPS uses iterative normalization and lookback to estimate the average reward and achieves a better bias-variance trade-off than IPS. In addition to high variance, violation of the Unconfoundedness assumption is another challenge of utilizing IPS in RS. That is, the treatment mechanism is \emph{identifiable}~\cite{glymour2016causal, mohan2021graphical} from observed covariates due to the existence of unobserved ones, which leads to the inaccurate estimate of propensity score and the disagreement between the online and offline evaluations. To address the uncertainty brought by the identifiability issue, ~\cite{xu2020adversarial} proposes minimax empirical risk formulation, which can be converted to an adversarial game between two recommendation models via duality arguments and relaxations.

More recently, Liu et al.~\cite{liu2022practical} propose Regularized per-item IPS (RIIPS) with an additional penalty function that constrains the difference in recommended outcomes between the deployed system and the new system so that the explosion of propensity scores can be avoided.

\subsubsection{Doubly Robust}
Doubly Robust (DR)~\cite{funk2011doubly, dudik2014doubly, jiang2016doubly, wang2019doubly} is another powerful and effective causal method account for the MNAR issue. To understand DR, let us consider the two common-used approaches to mitigate against MNAR: direct method (DM)~\cite{beygelzimer2009offset} and IPS~\cite{saito2021evaluating}. The former designs a model (linear regression, deep neural network, etc.) to directly learn the missing outcomes based on the observed data, which has low variance due to the advantage of supervised learning but suffers from high bias caused by unmet IID assumptions, denoted as~\cite{saito2021evaluating}:
\begin{equation}
\hat{R}_{\mathrm{DM}}\left(\pi_{0} ; \mathcal{O}_{\pi_{0}}, \hat{y}\left(x_{k}, t\right) \right):=\frac{1}{|\mathcal{O}_{\pi_{0}}|} \sum_{k=1}^{|\mathcal{O}_{\pi_{0}}|}\textrm{Pr}_{\pi}\left(t=1 \mid x_{k}\right) \hat{y}\left(x_{k}, t\right),
\end{equation}
where $\hat{y}\left(x, t\right)$ is the estimated outcomes. The latter, though unbiased theoretically, often causes training losses to oscillate stemming from the inverse of propensity with high variance~\cite{thomas2016data}. What DR does is to combine the direct method and IPS, which takes advantage of both and overcomes their limitations:
\begin{equation}
\hat{R}_{\mathrm{DR}}\left(\pi ; \mathcal{O}_{\pi_0}, \hat{r}\right) := \hat{R}_{\mathrm{DM}}\left(\pi ; \mathcal{O}_{\pi_0}, \hat{y}\left(x_{k}, t\right) \right) + 
\frac{1}{|\mathcal{O}_{\pi_{0}}|} \sum_{k=1}^{|\mathcal{O}_{\pi_{0}}|} 
\frac{e_{\pi}(X)}{e_{\pi_0}(X)} \left(y_{k}-\hat{y}\left(x_{k}, t_{k}\right)\right). 
\end{equation}
DR uses the estimated outcomes to decrease the variance of IPS. It is also \emph{doubly robust} in that it is consistent with the policy reward value if either the propensity scores or the imputed outcomes are accurate for all user-item pairs~\cite{wang2019doubly, saito2021evaluating}. By the way, advanced versions like Switch-DR~\cite{wang2017optimal} and DRos (Doubly Robust with Optimistic Shrinkage)~\cite{su2020doubly} are proposed to further control the variance.

Based on the above advantages, DR has found an increasingly wide utilization in RSs~\cite{yuan2019improving, wang2019doubly, zhang2020large, guo2021enhanced, kiyohara2022doubly, mondal2022aspire, xiao2022towards, dai2022generalized, song2023cdr, li2023should}. Wang et al.~\cite{wang2019doubly} utilize DR for unbiased RS prediction and further propose a joint learning approach that simultaneously learns rating prediction and propensity to guarantee a low prediction inaccuracy at inference time. Yuan et al.~\cite{yuan2019improving} propose a propensity-free doubly robust method to address the issue that samples with low propensity scores are absent in the observed dataset. Zhang et al.~\cite{zhang2020large} propose Multi-DR based on a multi-task learning framework to address selection bias and data sparsity issues in CVR estimation. Gun et al.~\cite{guo2021enhanced} propose the MRDR (more robust doubly robust) estimator to further reduce the variance caused by inaccurate imputed outcomes in DR while retaining its double robustness. In addition, Kiyohara et al.~\cite{kiyohara2022doubly} expand previous RIPS to Cascade Doubly Robust estimator, which has the same user interaction assumption as RIPS. Xiao et al.~\cite{xiao2022towards} propose an information bottleneck-based approach to effectively learn the DR estimator for the estimation of recommendation uplift, with the hope of a better trade-off between the bias and variance of propensity scores. Dai et al.~\cite{dai2022generalized} learns imputation with balancing the variance and bias of DR loss. More recently, Song et al.~\cite{song2023cdr} filter imputation data through examination of their mean and variance, in order to reduce poisonous imputations that significantly deviate from the truth and impair the debiasing performance.

\subsection{Causal Effect Strategy}

The most critical and fundamental role of causal inference is to estimate the causal effects from observational data, which has a variety of applications in real-world recommender systems. Some works are dedicated to estimating and enhancing the treatment effect of a recommender policy on specific customer outcomes, namely uplift~\cite{gutierrez2017causal}. In such scenarios, the causal effect is typically implemented as either a direct or indirect optimization goal, aiming to maximize platform benefits. Additionally, treatment effects extend to other application areas in recommender systems, serving purposes beyond uplift.

It is crucial to highlight that within the PO framework, the causal relationships between variables are not the focal point while calculating causal effect, and all variables affecting potential outcomes except treatment will be treated as covariates.

\begin{table*}[pt]
  \small
  \centering
  \vspace{-3mm}
  \caption{\normalsize Summary of causal effect strategies for recommendation.}
  \vspace{-3mm}
  \setlength{\tabcolsep}{2.4mm}{
    \begin{tabular}{c|c|c|c|c|c}
    \toprule
    \textbf{Category} & \textbf{Model} & \textbf{Causal method} & \textbf{Backbone model} & \textbf{Issue of concern} & \textbf{Year} \\
    \midrule
    \multirow{5}[2]{*}{\makecell{Causal \\effect\\ for\\ Uplift}} & ULRMF, ULBPR~\cite{sato2019uplift} & IPS, SNIPS, ATE & MF  & \multirow{5}[2]{*}{Uplift} & 2019 \\
          & ~\cite{goldenberg2020free}  & CATE  & Xgboost~\cite{chen2016xgboost} &       & 2020 \\
          & AUUC-max~\cite{betlei2021uplift} & CATE  & \makecell{Linear\\/Wide \& Deep\\  }  &       & 2021 \\
          & CausCF~\cite{xie2021causcf} & CATE  & MF  &       & 2021 \\
          & ASPIRE~\cite{mondal2022aspire} & DR, ATE & LightGBM~\cite{ke2017lightgbm} &       & 2022 \\
    \midrule
    \multirow{6}[2]{*}{\makecell{Causal \\effect\\ beyond\\ Uplift}} & ~\cite{rosenfeld2017predicting} & ITE   & \makecell{Linear/regularized \\kernel methods} & Domain adaptation & 2017 \\
          & CausE~\cite{bonner2018causal} & ITE   & MF  & Domain adaptation & 2018 \\
          & ~\cite{mehrotra2020inferring} & TE    & \makecell{Structural \\state-space model~\cite{brodersen2015inferring}} & \makecell{Causal effect of\\a new track release} & 2020 \\
          & CACF~\cite{zhang2021causally} & ITE   & (custom-designed) & Unobserved confounding bias & 2021 \\
          & MCRec~\cite{yao2022device} & CATE  & DIN~\cite{zhou2018deep} & \makecell{Device-cloud \\recommendation} & 2022 \\
          & LRIR~\cite{tran2022most} & ITE, ATE & (custom-designed) & Disability employment & 2022 \\
    \bottomrule
    \end{tabular}}%
  \label{tab:effect}%
  \vspace{-6mm}
\end{table*}%

\subsubsection{Causal Effect for Uplift}

Uplift, denoting the causal effect of recommendations, refers to the increase in user interactions purely caused by recommendations. Typical evaluations of recommender systems regard positive user interactions as a success. However, a subset of these interactions might persist even in the absence of recommendations. This assertion is substantiated by the conclusion of Sharma et al.~\cite{sharma2015estimating}, which indicates that more than 75\% of click-throughs would still occur in the absence of recommendations. For marketing campaigns where Return on Investment (ROI) is paramount, targeting 'voluntary buyers' — individuals who would interact with or without any recommendations — is deemed unnecessary. Therefore, the industry regards uplift as a valuable metric for recommendations in expectation of higher rewards.



It is a natural application to introduce the causality concepts such as ATE and CATE for uplift modeling since the definition of uplift is a counterfactual problem and consistent with the objective of causal effect estimation~\cite{yamane2018uplift, zhang2021unified, gutierrez2017causal}. Causal approaches with traditional machine learning methods for uplift estimation include two-model approach~\cite{radcliffe2007using, nassif2013uplift}, transformed outcome~\cite{jaskowski2012uplift} and uplift trees~\cite{radcliffe2011real, rzepakowski2012decision}. Regarding recommender systems, uplift estimation on online A/B testing suffers from the high expense and large fluctuations due to user self-selection bias~\cite{sato2021online}, while uplift estimated offline is bedeviled by a wide variety of biases that could lead to MNAR. In order to deal with these issues, much of the literature has been published. Sato et al.~\cite{sato2019uplift} utilize SNIPS-based ATE to accomplish offline uplift-based evaluation. Goldenberg et al.~\cite{goldenberg2020free} leverage the Retrospective Estimation technique that relies solely on data with positive outcomes for CATE-based uplift modeling, which makes it especially suited for many recommendation scenarios where only the treatment outcomes are observable. ~\cite{betlei2021uplift} learns a model that directly optimizes an upper bound on AUUC, a popular uplift metric based on the uplift curves and unified with ATE~\cite{yamane2018uplift}. In addition,  CausCF~\cite{xie2021causcf} extends the classical MF to the tensor factorization with three dimensions—user, item, and treatment effect for better uplift performance. CF-MTL~\cite{li2023should} accounts for whether users actively accept the treatment, leading to a more granular classification of users, and then estimates the probability for each user type within a multi-task learning framework. It is worth mentioning that in the uplift modeling literature~\cite{diemert2018large,gutierrez2017causal, zhang2021unified}, there are two closely related metrics for uplift modeling, uplift and Qini curves, the latter of which is evaluated based on the ranking of conditional treatment effect estimations.

\subsubsection{Causal Effect beyond Uplift}
There are some other impressive recommendation works with causal effect~\cite{mehrotra2020inferring, zhang2021causally, rosenfeld2017predicting, bonner2018causal, yao2022device, tran2022most}. For example, ~\cite{mehrotra2020inferring} adapts a Bayesian model to infer the causal impact of new track releases, which may be an essential consideration in the design of music recommendation platforms. ~\cite{zhang2021causally} minimizes the distance between the traditional attention weights in the recommendation method and the ITE to reflect the true impact of the features on the interactions. ~\cite{rosenfeld2017predicting} and ~\cite{bonner2018causal} frames causal inference as a domain adaptation problem and leverages ITE with a large sample of biased data and a small sample of unbiased data to eliminate the bias problems, which are described in more detail in \ref{sec:domain_adaption}.


\subsection{Why Potential Outcomes Framework?}

\begin{figure}[t!]
	\centering
    \vspace{-4mm}
	\subfloat[The real world]{
		\centering
		\includegraphics[width=0.35\textheight, trim=0 5 0 0, clip]{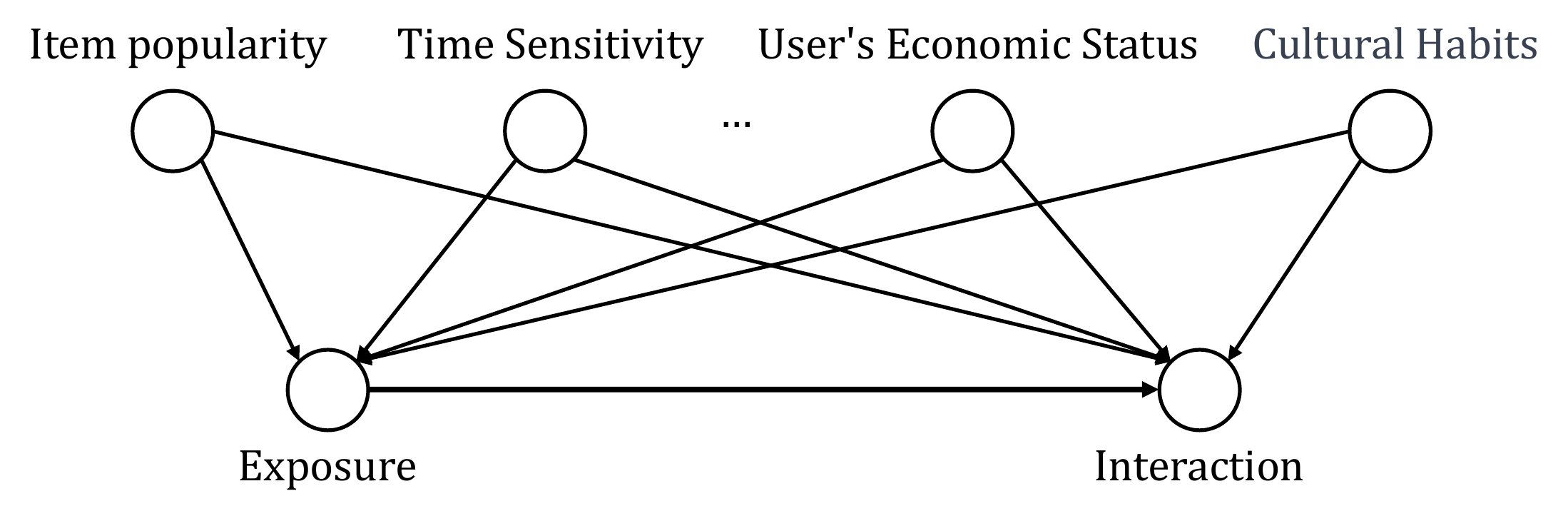}
	}
	\centering
	\subfloat[The world simulated by a randomized controlled trial]{
		\centering
		\includegraphics[width=0.35\textheight, trim=0 0 0 5, clip]{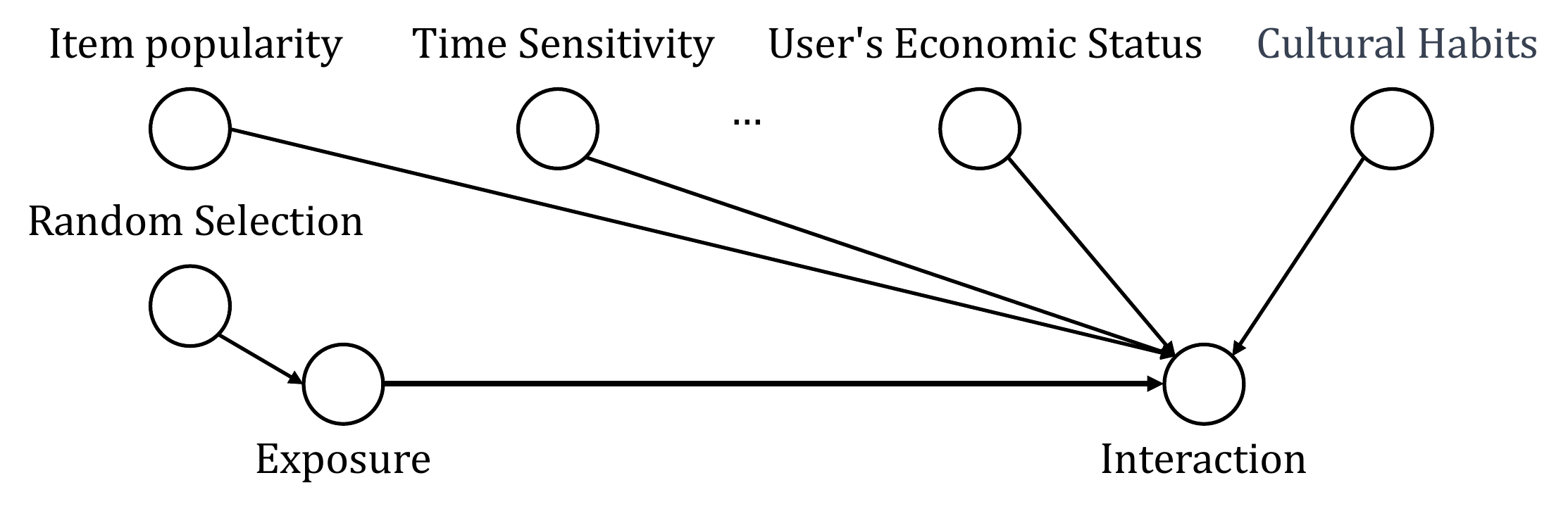}
	}
	\centering
    \vspace{-3mm}
	\caption{(a) Confounders such as item popularity and user's economic status influence both the likelihood of an item being recommended by the system (i.e., the treatment) and the user's final interaction (i.e., the potential outcome). (b) An ideal randomized controlled trial (RCT) disables the effect of confounders on the receipt of treatment, allowing for an accurate estimation of the desired treatment effect.}
    \vspace{-6mm}
	\label{fig:abtest}
\end{figure}

The PO framework has maintained its popularity in the realm of recommender systems since its inception due to its close association with A/B testing. Online A/B testing evaluates the performance of two different recommender policies through randomized experiments. Specifically, a user pool on the platform is randomly divided into a treatment group and a control group, with each group being exposed to one of the policies~\cite{gilotte2018offline}. Upon completion of the experiment, metrics such as revenue and click-through rates are compared to determine the policy to be adopted for future use. As illustrated in Fig.  \ref{fig:abtest}, the efficacy of A/B testing stems from the ideal randomized controlled trial (RCT) that disables all the confounders simultaneously affecting the treatment and the outcomes, thereby leading to a pure assessment of the policy's treatment effect on potential outcomes~\cite{pearl2009causality}. In practice, however, A/B testing often fails to achieve ideal randomization due to issues such as insufficient sample sizes leading to distributions that do not match the overall population. In such cases, methods like IPS from the PO framework can adjust sample weights, thus mitigating selection biases.

Due to the time-intensive and costly nature of online A/B testing, offline A/B testing serves as a more expedient and cost-effective approach to estimate the efficacy of recommended policies. Offline data are accumulated using the current recommendation system, referred to as the logging policy; hence, we cannot use the direct estimation of the target policy on offline data since it was not collected under the conditions of the target policy. Instead, it is necessary to re-weight the importance of samples to align with the data distribution that would be expected under the target policy. Moreover, in the context of marketing campaigns, beyond accurately estimating the effect of a policy, it is crucial to calculate each user's uplift to precisely identify the target users. CATE-based uplift modeling is adept at distinguishing between voluntary buyers and the persuadables—who would only interact in reaction to an incentive—thus fulfilling marketing objectives.


The policy evaluation methods mentioned above can also be transformed into the optimization objectives (e.g., the loss function) of recommendation algorithms, thereby aligning the model's optimization targets with evaluation goals. 

Policy evaluation is fundamentally crucial for several reasons: (1) its sustained significance since the inception of recommender systems; (2) its steadfastness amidst the development of recommendation algorithm technology shift; and (3) its archetypal alignment with the PO framework as a typical counterfactual question, where the true outcome of a unit under an alternative treatment remains perpetually indeterminate. Therefore, the PO framework seizes a prominent stage for deployment within recommender systems.

\section{SCM-based Methods}
\label{sec:scm_based}

Unlike the PO framework, Structural Causal Model explicitly expresses the causal relationship between variables on a causal graph, based on the experiences, before analyzing the causal effect. Its intuitive features make it win undivided admiration among researchers in computer field. In this section, the corresponding strategies is classified according to their causal structures, i.e., collider, mediator, and confounder. We focus on how researchers abstract recommendation issues into causal problems with causal graphs and exploit tools in causal inference to cope with it.

\subsection{Causal Recommendation with Collider Structure}

As represented in Fig. ~\ref{fig:junction}(c), a collider node occurs when it receives effects from two or more other factors. Collider exists in recommender systems. For instance, item positions in the ranking list are influenced by user preference and item popularity. 

Analyzing the dependency between variables in collider structures will contribute to its utilization in recommender systems. Although $A$ and $C$ are independent, i.e., for all $a$ and $c$, $\textrm{Pr}(A=a|B=b)=\textrm{Pr}(A=a)$, conditioning on the collision node $C$ produces a dependence between the node’s parents, i.e., for some $a,b,c$, $\textrm{Pr}(A=a|B=b, C=c)=\textrm{Pr}(A=a|C=c)$.  
To understand the point, let us consider the most basic example where $C=A+B$, and 
$A$ and $B$ are independent variables~\cite{glymour2016causal}. In this case, given $C=10$, knowing $A=3$ means we can immediately calculate that $B=7$. Thus, $A$ and $B$ are dependent, given that $C=10$. This characteristic inspires us that in RS issues with collider structure, knowing the common effect and one of the causes would provide information for another effect ~\cite{zhang2022causal}.

\begin{figure}[t!]
	\centering
    \vspace{-4mm}
	\subfloat[DICE]{
		\centering
		\includegraphics[width=0.23\textwidth, trim=0 -20 0 -10, clip]{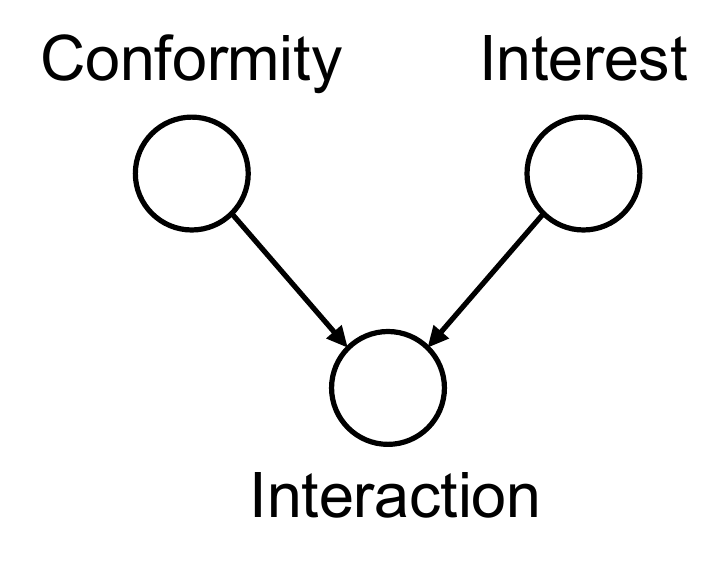}
	}
	\centering
	\subfloat[CIGC]{
		\centering
		\includegraphics[width=0.28\textwidth, trim=8 0 10 5, clip]{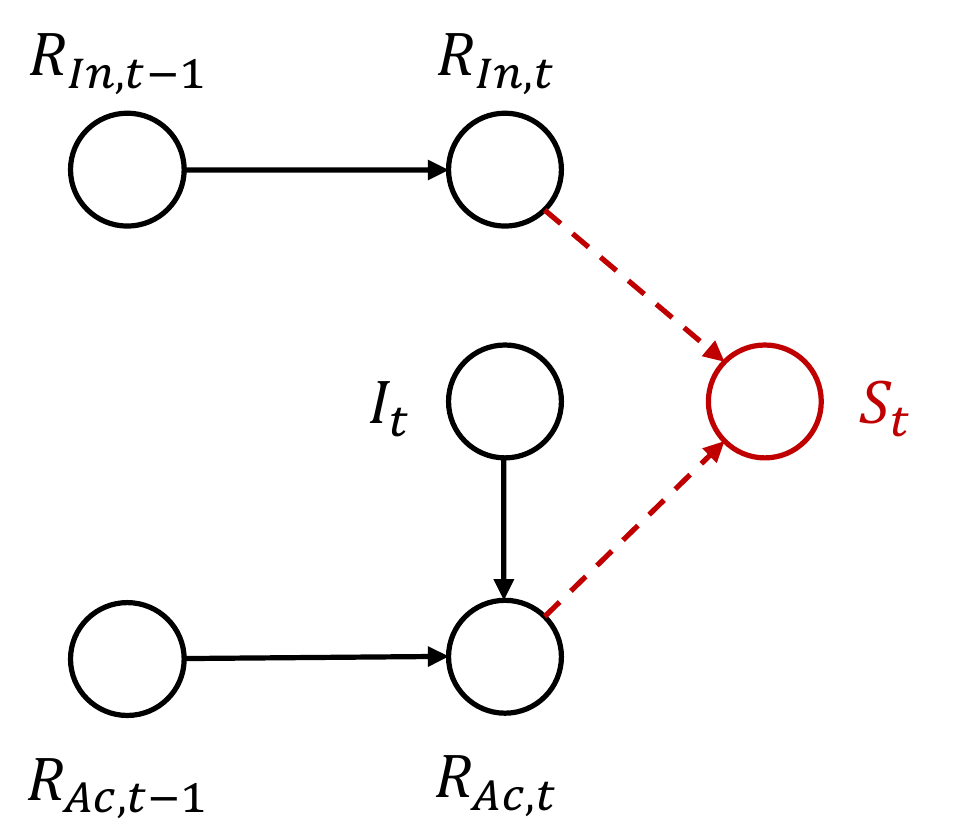}
	}
	\centering
    \vspace{-3mm}
	\caption{Causal graphs of collider in recommender systems.}
    \vspace{-6mm}
	\label{fig:collider}
\end{figure}

Though collider structures permeate RSs, they are usually compounded by other causal relationships and are treated as other causal structures, which results in minor literature discussing purely colliders. A representative work is DICE~\cite{zheng2021disentangling}, which is proposed by Zheng et al. and tracks the popularity issue from the user’s perspective instead of eliminating popularity bias from the item’s perspective. Zheng et al. argue that users’ interactions are driven by individual interest as well as users’ conformity, which is independent of user interest and describes how users tend to follow other people, and provides a causal graph as shown in Fig. \ref{fig:collider} (a). From this point of view,  DICE splits user and item embeddings into interest and conformity embeddings, respectively, and learns disentangled representations with conformity-specific and interest-specific data, driven by the colliding effect: if a user interacts with a less popular item, not conforming to the mainstream, it usually indicates that the user is highly interested in the item itself, and vice versa. Further, ~\cite{ding2022causal} proposes CIGC (Causal Incremental Graph Convolution), which includes a new operator named CED (Colliding Effect Distillation), to efficiently retrain graph convolution network (GCN) based recommender models. CED frames the whole incremental training phase as a causal graph (see Fig. \ref{fig:collider} (b)) and create a collider $S_t$ between inactive nodes $R_{In,t}$ and new data $R_{Ac,t}$, which is represented as the pair-wise distance. Therefore, the incremental integration data $I_t $ can update both $R_{Ac,t}$ and $R_{In,t} $, since conditioning on the collider $S_t$ opens the path $I_t \rightarrow R_{Ac,t} \leftrightarrow  R_{In,t} $.

\subsection{Causal Recommendation with Mediator Structure}

When one variable causes another, it may not do it directly but through a set of mediating variables instead. For example, an item purchased by your friends increases your purchase probability not only directly through the recommendation that integrates social network, but also indirectly through increased trust in the item.

\begin{figure}[t!]
	\centering
    \vspace{-3mm}
	\subfloat[Simple mediator]{
		\centering
		\includegraphics[width=0.23\textwidth, trim=-20 -20 -20 0, clip]{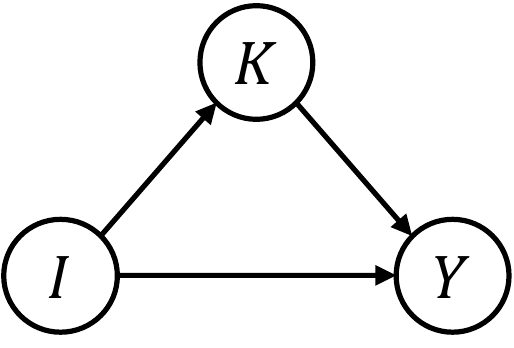}
	}
	\centering
	\subfloat[Mediator with confounder]{
		\centering
		\includegraphics[width=0.23\textwidth, trim=-20 0 -20 0, clip]{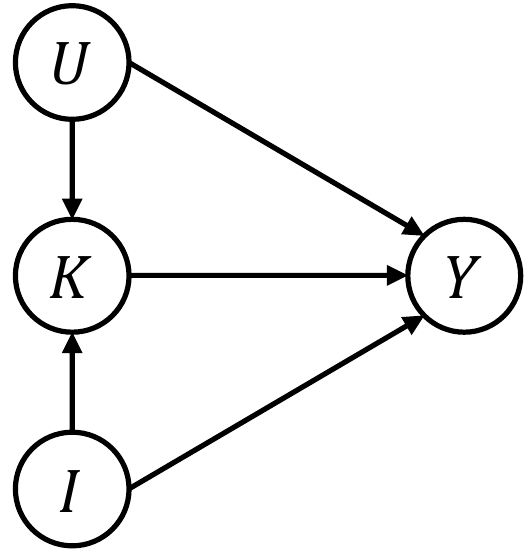}
	}
	\centering
    \vspace{-1mm}
	\caption{Causal graphs of mediators in recommender systems, where $I$, $Y$, $K$ and $U$ denote cause, effect, mediator and confounder variable of the mediator and the outcome, respectively.}
    \vspace{-2mm}
	\label{fig:mediator}
\end{figure}

The distinction between direct and indirect effects of the change of treatment on outcome is key to the utilization of the mediator structure, which can be done by conditioning on the mediating variable traditionally~\cite{glymour2016causal}. Specifically, as illustrated in Fig. \ref{fig:mediator} (a), the \textit{total effect} (ToE) of $I=i$ on $Y$ is defined as:
\begin{equation}
ToE=Y(I=i, K(I=i))-Y(I=i^*, K(I=i^*)),
\end{equation}
$I = i^{*}$ refers to the situation where the value of $I$ is different from the reality, i.e., counterfactual.
Total effect can be further decomposed into \textit{natural direct effect} (NDE) and \textit{total indirect effect} (TIE). NDE reflects the effect of $I$ on $Y$ through the direct path, i.e., $I \rightarrow Y$, while $K$ is set to the value when $I=i^*$:
\begin{equation}
    NDE = Y(I=i, K(I=i^*))-Y(I=i^*, K(I=i^*)).
\end{equation}
TIE is defined as the difference between TE and NIE, denoted as:
\begin{equation}
\label{eq:tie}
    TIE = ToE - NDE  = Y(I=i, K(I=i))-Y(I=i, K(I=i^*)),
\end{equation}
which represents the effect of $I$ on $Y$ through the indirect path $I \rightarrow K \rightarrow Y$. TE can also be decomposed into \textit{natural indirect effect} (NIE) and \textit{total direct effect} (TDE). NIE represents the effect of $I$ on $Y$ through the mediator, i.e., $I \rightarrow K \rightarrow  Y$, while the direct effect on $I \rightarrow Y$ is blocked by setting $I$ as $I*$, denoted as:
\begin{equation}
    NIE = Y(I=i*, K(I=i))-Y(I=i^*, K(I=i^*)).
\end{equation}
In linear systems, NIE and TIE have the same value, and NDE and TDE have the same value~\cite{glymour2016causal, pearl2022direct}.

However, if there are confounders of the mediator and the outcome, as the case of ~\cite{wei2021model} shown in Fig. \ref{fig:mediator} (b), conditioning on the mediator means conditioning on a collider, and thus indirect dependence will pass through the confounder to the outcome and misguide the calculation of indirect effect. To tackle the problem, we should intervene on the mediator, which involves counterfactuals. The controlled direct effect (CDE) on $Y$ of $I$ is defined as:
\begin{equation}
    CDE = Y(do(I=i), do(K=k))-Y(do(I=i^*), do(K=k)).
\end{equation}
The difference between NDE and CDE is explained in ~\cite{glymour2016causal}.

\begin{table*}[pt]
  \small
  \centering
  \vspace{-3mm}
  \caption{\normalsize Summary of recommendation models with collider structure and mediator structure.}
  \vspace{-3mm}
  \setlength{\tabcolsep}{2.2mm}{
    \begin{tabular}{c|c|c|c|c|c}
    \toprule
    \textbf{Category} & \textbf{Model} & \textbf{Causal method} & \textbf{Backbone model} & \textbf{Issue of concern} & \textbf{Year} \\
    \midrule
    \multirow{2}[2]{*}{\makecell{Causal\\ recommendation\\ with collider structure}} & DICE~\cite{zheng2021disentangling} & \makecell{(causal view)} & MF(multi-task) & Popularity bias & 2021 \\
          & CIGC~\cite{ding2022causal} & \makecell{Intervention on \\the cause factor} & LightGCN~\cite{he2020lightgcn} & GCN model retraining & 2022 \\
    \midrule
    \multirow{6}[2]{*}{\makecell{Causal\\ recommendation \\with mediator \\structure}} & ~\cite{choi2011influence} & Mediation analysis & -     & Effect of social presence & 2011 \\
          & ~\cite{luo2013impact} & Mediation analysis & -     & Effect of informational factors & 2013 \\
          & CMA~\cite{yin2019identification} & NDE, TIE & -     & Effect of induced change & 2019 \\
          & MACR~\cite{wei2021model} & TIE   & (model-agnostic, multi-task) & Popularity bias & 2021 \\
          & CIRS~\cite{gao2022cirs} & \makecell{Intervention on \\the mediator} & PPO~\cite{schulman2017proximal} & Filter bubble~\cite{pariser2011filter} & 2022 \\
          & CCF~\cite{xu2021causal} & \makecell{Intervention on\\ the mediator, \\counterfactuals} & \makecell{NCF~\cite{he2017neural}, \\GRU4Rec~\cite{hidasi2015session}, etc.} & Historical bias & 2023 \\
    \bottomrule
    \end{tabular}}%
  \vspace{-3mm}
  \label{tab:collider_mediator}%
\end{table*}%

Some works are generally interested in how much of the treatment’s causal effect on variable $Y$ is direct and how much is indirect, which is usually explored with the technique of \textit{mediation analysis}~\cite{kenny1979correlation, baron1986moderator}, similar to SCM but without exogenous variables and the introduction of counterfactuals. For example, in early studies, ~\cite{choi2011influence} conducts an experiment varying the level of social presence over hundreds of testers and examines the effect of social presence on users’ reuse intention and trust through mediation analysis. A similar structure is used to evaluate how electronic word-of-mouth affects user interactions in ~\cite{luo2013impact}. Further, Yin et al.~\cite{yin2019identification} aim to separate the direct effects of the change in user behaviors in the tested product from the effect of changes in user behaviors in other products, aka induced changes, for example, the effects of significant lifts in CTR on the recommendation list and of significant decreases in CTR on organic search results on the final insignificant lifts in the sitewide CVR during the A/B test of a new version of recommendation module. Therefore, they use causal mediation analysis (CMA) of potential outcome framework to estimate causal effects of the induced changes and also discuss the estimation under the situation that multiple unmeasured causally-dependent mediators exist with the help of a directed acyclic graph.

Some other works utilize Pearl’s counterfactual tool to cope with mediator structure in order to improve accuracy~\cite{wei2021model, xu2021causal, gao2022cirs}. Wei et al.~\cite{wei2021model} explore the popularity issue with the SCM framework and formulate the causal graph as Fig. \ref{fig:mediator} (b) shown, in which the probability of interaction $Y$ is influenced by three main factors: user-item matching ($K(U, I) \rightarrow Y$), item popularity ($I \rightarrow Y$) and user conformity ($U \rightarrow Y$), the last two of which are usually ignored by existing models and thus result in the terrible Matthew effect. Following this causal graph, Wei et al. propose MACR (Model-Agnostic Counterfactual Reasoning), a multi-task framework that consists of three modules to jointly learn the effects of $U \rightarrow Y$, $U\&I \rightarrow K \rightarrow Y$, and $I \rightarrow Y$,  respectively during recommender training and estimates TIE of $I$ on $Y$ in counterfactual inference:
\begin{equation}
\begin{aligned}
\label{eq:tie_macr}
 TIE= &ToE - NDE  \\
= &Y(U=u, I=i, K=K(U=u, I=i))-Y(U=u, I=i, do(K=K(U=u^*,I=i^*))) 	\\
= &Y_k(K(U=u, I=i))*Y_u(U=u)*Y_i(I=i)-Y_k(K(U=u^*, I=i^*)) *Y_u(U=u)*Y_i(I=i)\\
=&\hat{y}_{k} * \sigma\left(\hat{y}_{i}\right) * \sigma\left(\hat{y}_{u}\right)-c * \sigma\left(\hat{y}_{i}\right) * \sigma\left(\hat{y}_{u}\right),
\end{aligned}
\end{equation}
where $\sigma(\cdot)$ denotes the sigmoid function, and $c$ is a hyper-parameter that represents $Y_k(K(U=u*, I=I*))$, the reference situation of $Y_k(K(U=u, I=i))$. With counterfactual inference, MACR could rank items without popularity bias by reducing the direct effect from item properties to the ranking score.

\begin{figure}[t!]
	\centering
    \vspace{-3mm}
	\subfloat[]{
		\centering
		\includegraphics[width=0.23\textwidth, trim=-20 0 -20 0, clip]{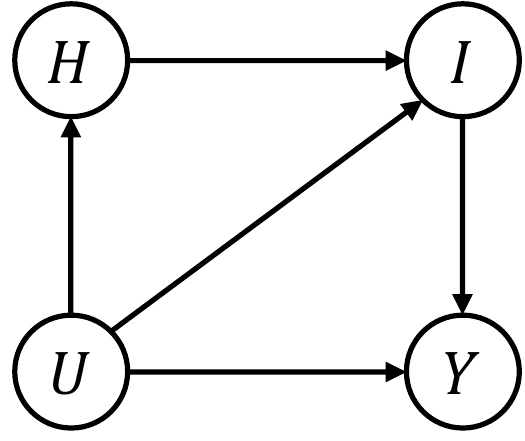}
	}
	\centering
	\subfloat[]{
		\centering
		\includegraphics[width=0.23\textwidth, trim=-20 0 -20 0, clip]{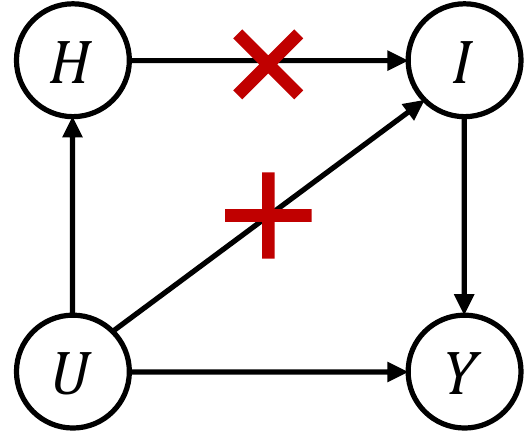}
	}
	\centering
    \vspace{-2mm}
	\caption{(a-b) Causal graphs of the CCF model before and after intervention. $U$ and $I$ are user and item representation, respectively, $Y$ is preference score, and $H$ denotes user interaction history.}
    \vspace{-2mm}
	\label{fig:mediator_ccf}
\end{figure}

The work by Xu et al.~\cite{xu2021causal} regards the user interaction history $H$ as a mediator (Fig. \ref{fig:mediator_ccf} (a)) and proposes CCF (Causal Collaborative Filtering) to estimate $\textrm{Pr}(Y=y|U=u, do(I=i))$, where $u, i$ is a user-item pair and $y$ is the preference score for the pair. More specifically, $H=f_h(U=u)$ is a database retrieval operation that returns a user’s interaction history from the observational data, $I=f_0(U=u, H=h)$ means the recommended item $I$ returned from the already deployed recommendation system based on the user and the user’s interaction history, and $Y=f(U=u, I=i)$ represents the estimation of unbiased user preference on the item. $\textrm{Pr}(Y=y|U=u, do(I=i))$ adopts the conditional intervention to consider both observed and unobserved (counterfactual) interaction history, as presented in Fig. \ref{fig:mediator_ccf} (b). The derivation result of $\textrm{Pr}(Y=y|U=u,do(I=i))$ is given:
\begin{equation}
\begin{aligned}
\textrm{Pr}(Y=y \mid U=u,do(I=i)) 
\approx & \textrm{Pr}(Y=y \mid U=u,do(I=f_0(U=u, H=h))) \\
 = & \sum_{h} \textrm{Pr}(y \mid u, h, f_0(u, h))\textrm{Pr}(h\mid u)\end{aligned}
\end{equation}
It is tempting to conclude that if trained only with observed history $h$, f(U=u, I=i) would naturally degenerate to the original recommendation model $f_0(U=u, H=h)$. Therefore, Xu et al. adopt a heuristic-based approach to generate counterfactual history $h’$.

\subsection{Causal Recommendation with Confounder Structure}

There is a large volume of published studies investigating the confounding structures in recommendation since a lot of data biases widespread in recommender systems are, essentially, confounding biases mentioned in Section \ref{sec:MNAR}. Approaches to tackle confounder structures of existing literature can be categorized into four types: with back-door adjustment, with instrumental variables, with front-door adjustment, and with deep learning based intervention.

\subsubsection{The Back-door-based Approach}

\begin{figure}[t!]
	\centering
    \vspace{-3mm}
	\subfloat[]{
		\centering
		\includegraphics[height=0.12\textheight, trim=-5 0 -5 0, clip]{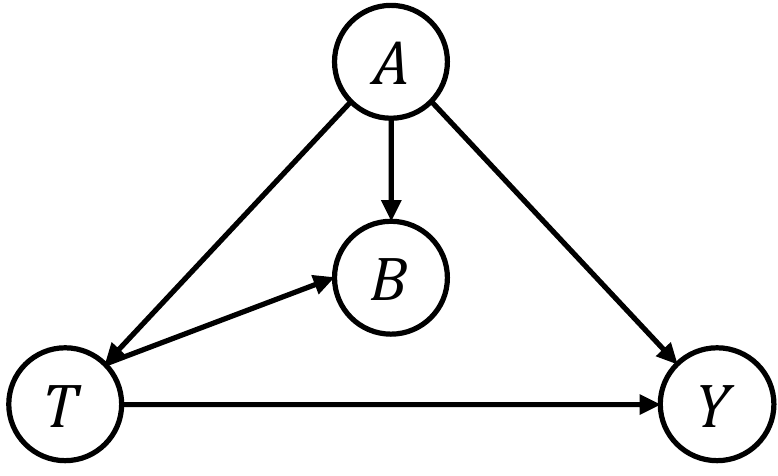}
	}
	\centering
	\subfloat[]{
		\centering
		\includegraphics[height=0.12\textheight, trim=-5 0 -5 0, clip]{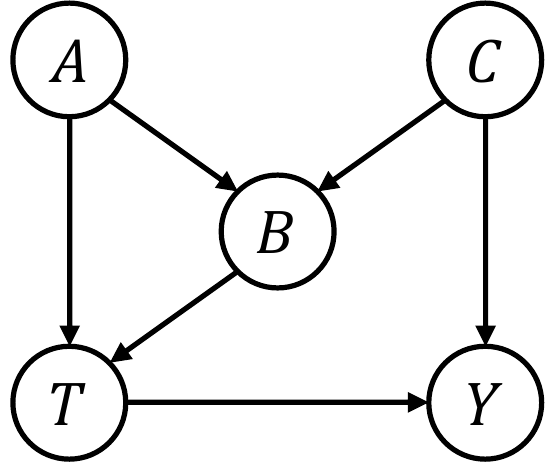}
	}
	\centering
    \vspace{-1mm}
	\caption{Causal graphs illustrating the back-door path.}
    \vspace{-4mm}
	\label{fig:confounder_back}
\end{figure}

Before introducing the back-door adjustment approaches, let us brieﬂy review the definitions of back-door path and back-door criterion~\cite{imbens2015causal}. 
\begin{definition}[Back-door Path]
\label{def:backdoor_path}
Given a pair of treatment $T$ and outcome variable $Y$, a path connecting $T$ and $Y$ is a back-door path for $(T, Y)$ if it satisfies that
\begin{enumerate}
\item it is not a directed path (it contains an arrow pointing into $T$); and
\item it is not blocked (it has no collider).
\end{enumerate}
\end{definition}
Back-door path help us to identify confounders, which is the central node of a fork on a back-door path of $(T, Y)$. The following two examples will help to illustrate it~\cite{pearl2018book}. In Fig. \ref{fig:confounder_back} (a), there is one back-door path from $T$ to $Y$, $T \leftarrow A \rightarrow Y$, indicating that $A$ is the confounder. For the estimation the effect of $T$ on $Y$, we should eliminate the confounding bias by either controlling $A$ to block the back-door path or running a randomized controlled experiment. Note that $T \rightarrow B \leftarrow A \rightarrow Y$ is blocked by the collider at $B$ and, therefore, not a back-door path. In Fig. \ref{fig:confounder_back} (b), we can control for $C$ to close the back-door path $T \leftarrow B \leftarrow C \rightarrow Y$. Here we present the formal definition of the back-door criterion to deal with the confounding effects.
\begin{definition}[Back-door Criterion]
\label{def:backdoor_criterion}
Given a pair of treatment $T$ and outcome variable $Y$, a set of variables $X$ satisfied the back-door criterion if $X$ blocks all back-door paths of $(T, Y)$.
\end{definition}

Based on the Back-door Criterion, we can further derive the Back-door Adjustment Theorem, which adjusts fewer variables compared to the Causal Effect Rule (Definition \ref{def:ce_rule}).

\begin{definition}[Back-door Adjustment]
\label{def:backdoor_adjustment}
If a set of variables $X$ satisfies the back-door criterion for $T$ and $Y$, the causal effect of $T$ on $Y$ is identifiable and given by the formula:
\begin{equation}
\textrm{Pr}(Y=y \mid do(T=t)) =\sum_{x} \textrm{Pr}(Y=y \mid T=t, X=x) \textrm{Pr}(X=x),
\end{equation}
\end{definition}

\begin{figure}[t!]
\centering
\vspace{-1mm}
\includegraphics[height=0.12\textheight,trim=0 0 0 0 ,clip]{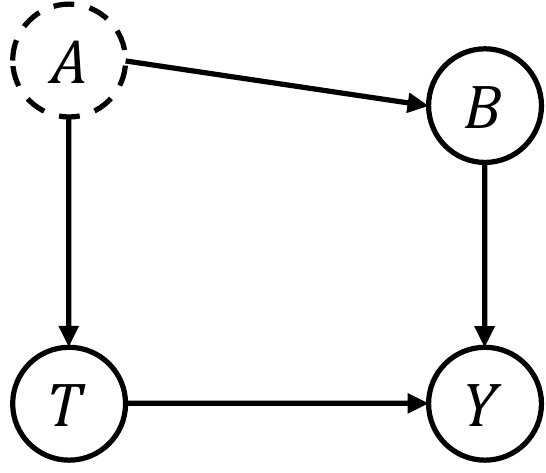} 
\vspace{-1mm}
\caption{A causal graph representing the relationship between recommendation ($T$), click ($Y$), consuming desire ($A$), and number of recent interactions $B$. The dotted circle indicates this variable is unobservable.}
\vspace{-1mm}
\label{fig:backdoor_ex}
\end{figure}

\begin{figure}[t!]
\centering
\vspace{-1mm}
\includegraphics[height=0.13\textheight,trim=0 0 0 0 ,clip]{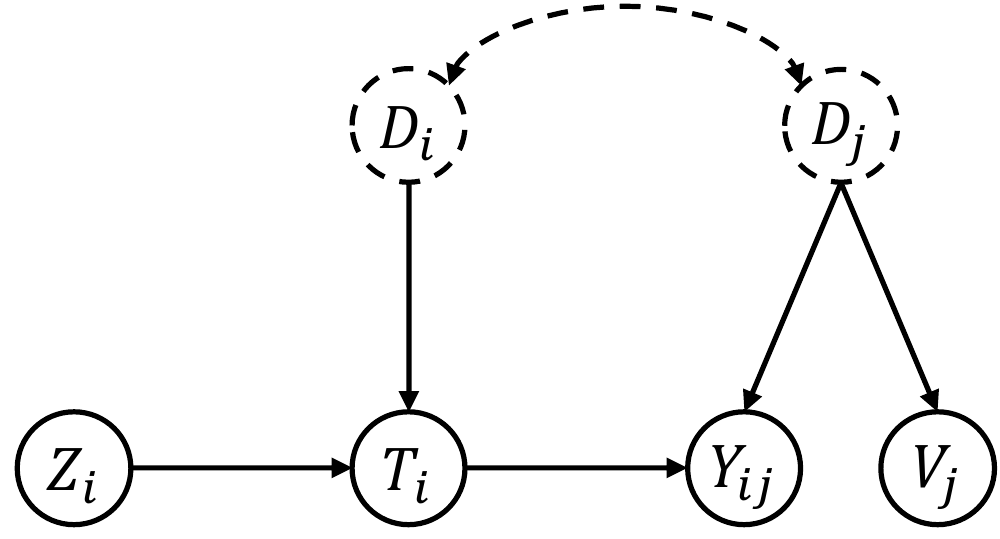} 
\vspace{-1mm}
\caption{A causal graph showing the relationships between a sudden shock in traffic 
($Z_{i}$), total exposure of the focal product $i$ ($T_{i}$), product demand $D_i$ and $D_j$, recommendation click-through of related product $j$ from the focal product $i$($Y_{ij}$). The focal product $i$ experiences an instantaneous shock $Z_{i}$ in traffic while the product $j$ recommended shown alongside does not. $V_j$ means direct exposure of product $j$ (e.g., through search or browsing), which is not influenced by recommendation.
}
\vspace{-3mm}
\label{fig:sharma2015estimating}
\end{figure}

To see what this means in practice, let us look at a concrete example, as presented in Fig \ref{fig:backdoor_ex}. Suppose we need to evaluate the effect of recommendation ($T$) on user’s click behavior ($Y$) of a newly deployed recommendation strategy on an online shopping platform. However, the time-varying consuming desire ($A$) makes it difficult to compare the effect with that of the existing one. For example, users might be more willing to spend due to the proximity of holidays, resulting in a seemly better recommendation effect of the tested policy. However, the consuming desire is unmeasurable for \textit{do}-calculation. Instead, we could control for an observed variable, the number of recent interactions $B$, that fits the back-door criterion from $T$ to $Y$. Therefore, adjusting for $B$ to block the back-door path $T \leftarrow A \rightarrow B \rightarrow  Y$ will give us the true causal effect of recommendation $T$ on click $Y$, formulated as:
\begin{equation}
\begin{aligned}
\textrm{Pr}(Y=y \mid do(T=t))=\sum_{x} \textrm{Pr}(Y=y \mid T=t, B=b) \textrm{Pr}(B=b).
\end{aligned}
\vspace{-2mm}
\end{equation}

Some literature on recommendation issues with confounder structures introduces the theory of back-door criterion~\cite{huang2012exploring, sharma2015estimating, tran2021recommending, wang2021clicks}. ~\cite{huang2012exploring} utilizes the back-door criterion to verify whether or not word-of-mouth recommendations can influence users’ evaluation of the recommended items. Sharma et al.~\cite{sharma2015estimating} treat an instantaneous shock in direct traffic as an \textit{instrumental variable} to answer the counterfactual question from purely observational data: how much interaction activity would there have been on the online shopping website if recommendations were absent, and apply the back-door criterion to block the possible unobserved confounding effect between the “exposure” $T_i$ and “click” $Y_{ij}$, as Fig. \ref{fig:sharma2015estimating} shown. Besides, Tran et al.~\cite{tran2021recommending} consider the job personal recommendation issue in Disability Employment Services and present a causality-based method to tackle the problem, in which the covariate set is determined by the back-door criterion.

\begin{figure}[t!]
	\centering
    \vspace{-3mm}
	\subfloat[DecRS]{
		\centering
		\includegraphics[height=0.14\textheight, trim=-15 -12 -15 -12, clip]{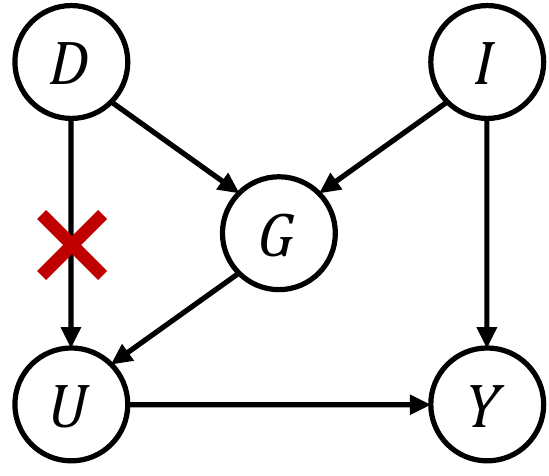}
	}
	\centering
	\subfloat[PDA]{
		\centering
		\includegraphics[height=0.14\textheight, trim=-15 0 -15 0, clip]{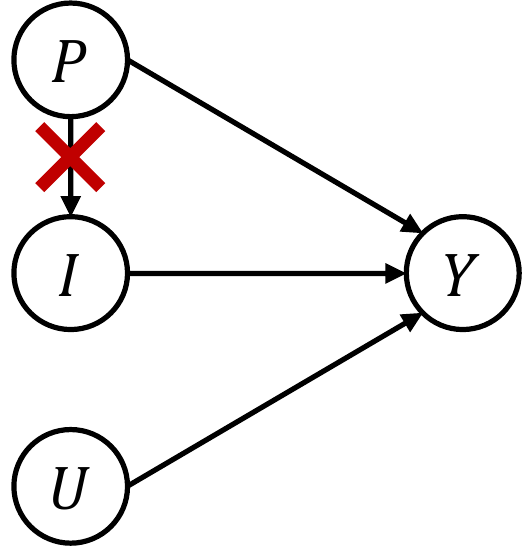}
	}
	\centering
    \vspace{-2mm}
	\caption{(a) The causal graph used in DecRS, where $U$ and $I$ denote user and item representation, $D$ represents the user historical distribution over item groups, $G$ is the group-level user representation, and $Y$ is the prediction score. (b) The causal graph of PDA, in which $U$ and $I$ denote user and item representation, $P$ is the item popularity, and $Y$ stands for user interactions.
}
    \vspace{-3mm}
	\label{fig:backpath}
\end{figure}

A multitude of studies employ back-door adjustment to block the back-door path by directly intervening on the treatment variable~\cite{wang2021deconfounded, zhang2021causal, he2022addressing, wang2022causal, zhan2022deconfounding, rajanala2022descover, xia2023user, zhang2023leveraging, yu2023causality, tsoumas2023evaluating}. For example, Wang et al.~\cite{wang2021deconfounded} propose the framework named DecRS (Deconfounded Recommender System) to eliminate bias amplification through intervention on the user representation $U$, which removes the effect of the historical user distribution over item groups $D$ on $U$, as Fig. \ref{fig:backpath} (a) shown. Zhang et al.~\cite{zhang2021causal} propose PDA (Popularity-bias Deconfounding and Adjusting) to eliminate the effect of item popularity $P$ through intervention on the item $I$ (see Fig. \ref{fig:backpath} (b)), denoted as:
\begin{equation}
\textrm{Pr}(Y=y \mid do(U=u, I=i))=\sum_{p} \textrm{Pr}(y \mid u, i, p) \textrm{Pr}(p \mid u,i)
=\textrm{Pr}(y \mid u, i, p) \textrm{Pr}(p),
\end{equation}
where $U$ denotes the user representation and $Y$ represents interactions. $\textrm{Pr}(y \mid u, i, p)$ and $\textrm{Pr}(p)$ are learned separately. It is worth mentioning that PDA can leverage popularity bias to enhance the recommendation performance by adjusting $\textrm{Pr}(p)$ in the inference stage, which can be regarded as counterfactual inference. More recently, Zhang et al.~\cite{zhang2023leveraging} address duration bias by identifying duration time as a confounder. Subsequently, they group data samples based on watch time feedback and craft novel duration supervision labels, thereby alleviating the confounding bias.

\begin{figure}[t!]
\centering
\vspace{-1mm}
\includegraphics[height=0.50\textheight,trim=0 10 0 0 ,clip]{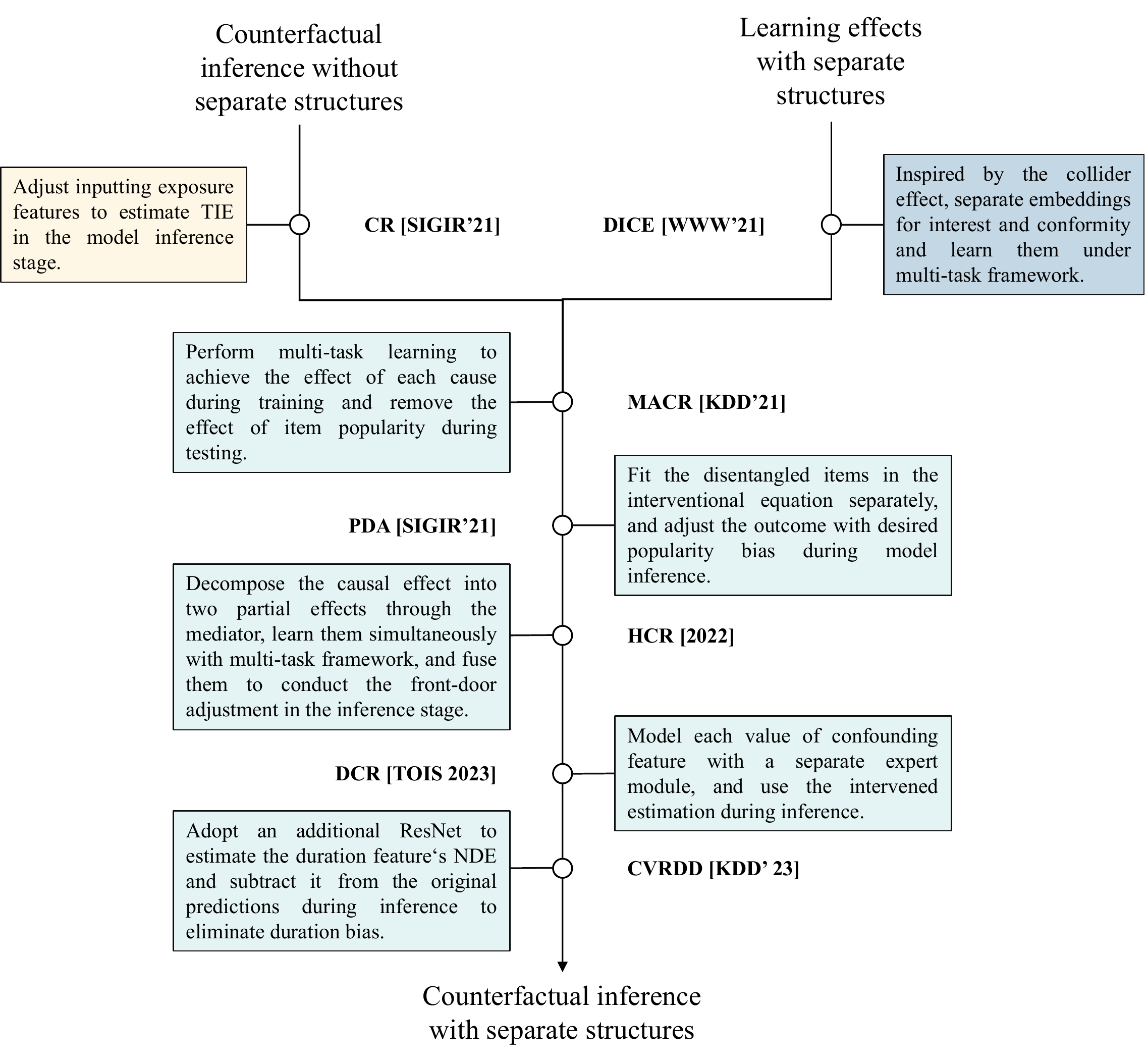} 
\vspace{-1mm}
\caption{Separate-learning-counterfactual-inference, a common pattern of SCM-based causal inference for recommender systems, learns causal effect with a separate structure or multi-task framework and performs counterfactual inference during testing.}
\vspace{-3mm}
\label{fig:he_type}
\end{figure}

In the above literature elaboration, we may find a series of works that accomplish the integration of SCM-based causal inference and recommender systems with a similar pattern, as shown in Fig. \ref{fig:he_type}: they first analyze the causal relationship between the variables regarding the concern issue and formulate the causal graph based on it; after theoretical analysis, a multi-task or separated structure is adopted to learn the causal effects of the variables on the potential outcome in the training phase; once the training has been completed, appropriate variables are selected to intervene during the inference stage, i.e., they are set to counterfactual values directly or indirectly, and the outcome is estimated based on applicable causal rules (e.g., backdoor adjustment, TIE, etc.) to conduct counterfactual inference. 



\subsubsection{Instrumental Variable-based Approach}
\begin{figure}[t!]
	\centering
    \vspace{-3mm}
	\subfloat[]{
		\centering
		\includegraphics[height=0.09\textheight,trim=0 0 0 0 ,clip]{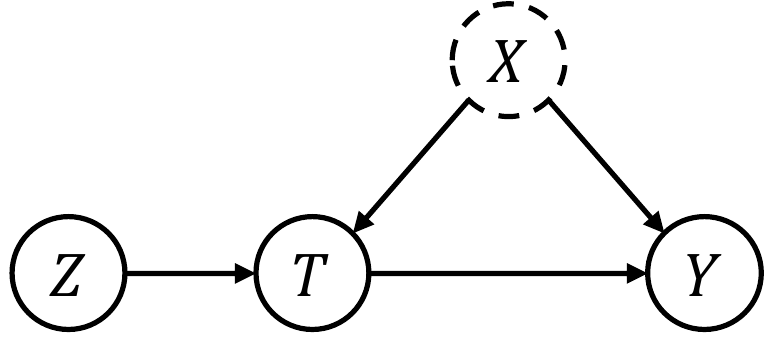}
	}
	\centering
	\subfloat[]{
		\centering
		\includegraphics[height=0.09\textheight,trim=0 0 0 0 ,clip]{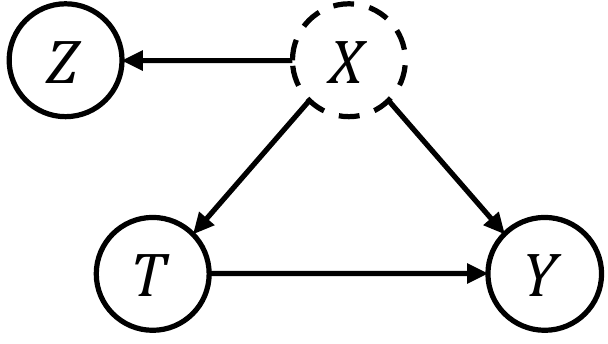}
	}
	\centering
    \vspace{-2mm}
	\caption{(a) The causal graph of a general setup for instrumental variables, where $Z$ is an instrumental variable. (b) Proxy variables is easier to be satisfied compared with the instrumental variables.
}
    \vspace{-3mm}
	\label{fig:iv_ex}
\end{figure}

The instrumental variable (IV) method is such a powerful approach for learning causal effects with confounders that it can be done even without controlling for, or collecting data on, the confounders~\cite{pearl2018book}. The instrumental variable causally influences the outcome only through the treatment (Fig. \ref{fig:iv_ex} (a)), defined as:
\begin{definition}[Instrumental Variable]
\label{def:iv}
Given an observed variable $Z$, covariates $X$, the treatment $T$ and the outcome $Y$, $Z$ is a valid instrumental variable (IV) for the causal effect of $T \rightarrow Y$ if $Z$ satisfies~\cite{angrist1996identification}: 
\begin{enumerate}
\item $Z \dep T \mid X$; and
\item $Z \ind Y \mid do(T), X$.
\end{enumerate}
\end{definition}
In practice, IV is often implemented in a two-stage lease squares (2SLS) procedure.

\begin{table*}[htbp]
  \centering
  \small  
  \vspace{-2mm}
  \caption{\normalsize Summary of recommendation models with confounder structure.}
  \vspace{-2mm}
  \setlength{\tabcolsep}{2.0mm}{
    \begin{tabular}{c|c|c|c|c}
    \toprule
    \textbf{Model} & \textbf{Causal method} & \textbf{Backbone model} & \textbf{Issue of concern} & \textbf{Year} \\
    \midrule
    ~\cite{huang2012exploring} & Back-door criterion & MF    & Effect of WOM recommendation & 2012 \\
    ~\cite{sharma2015estimating} & Back-door adjustment, IV & -     & Effect of recommendations & 2015 \\
    ~\cite{chaney2018algorithmic} & -     & MF, etc. & Feedback loop bias & 2018 \\
    DEMER~\cite{shang2019environment} & -     & \makecell{(RL)} & Unobserved confounding bias & 2019 \\
    CPR~\cite{yang2021top} & Back-door adjustment & (model-agnostic) & Data insufficiency & 2021 \\
    CauSeR~\cite{gupta2021causer}. & Back-door adjustment & SR-GNN~\cite{wu2019session} & Popularity bias in SBRSs & 2021 \\
    MCT~\cite{tran2021recommending} & Back-door criterion, CATE & (custom-designed) & Disability employment & 2021 \\
    DecRS~\cite{wang2021deconfounded} & Back-door adjustment & FM, NFM~\cite{he2017neuralfactorization} & Bias amplification & 2021 \\
    PDA~\cite{zhang2021causal} & Back-door adjustment & MF    & Popularity bias & 2021 \\
    CR~\cite{wang2021clicks} & Back-door criterion, TIE & \makecell{MMGCN~\cite{wei2019mmgcn}\\(multi-task)} & Clickbait & 2021 \\
    D2Q~\cite{zhan2022deconfounding} & Back-door adjustment & (custom-designed) & Duration bias & 2022 \\
    DeSCoVeR~\cite{rajanala2022descover} & Back-door adjustment & (custom-designed) & Venue recommendation & 2022 \\
    IV4Rec~\cite{si2022model} & IV    & DIN, NRHUB~\cite{wu2019neural} & Recommendation using search data & 2022 \\
    HCR~\cite{zhu2022mitigating} & Front-door adjustment & MMGCN & Unobserved confounding bias & 2022 \\
    DCR~\cite{he2022addressing} & Back-door adjustment & NFM   & \multicolumn{1}{c|}{Observed confounding bias} & 2023 \\
    CaDSI~\cite{wang2022causal} & Back-door adjustment & (custom-designed) & Observed confounding bias & 2023 \\
    DecUCB~\cite{xia2023user} & Back-door adjustment & (custom-designed, bandit) & Observed confounding bias & 2023 \\
    iDCF~\cite{zhang2023debiasing} & Proxy variable & MF & Unobserved confounding bias & 2023 \\
    CVRDD~\cite{tang2023counterfactual} & TIE & MLP(model-agnostic) & Duration bias & 2023 \\
    DML~\cite{zhang2023leveraging} & Back-door adjustment & MMoE & Duration bias & 2023 \\
    CGSR~\cite{yu2023causality} & Back-door adjustment & (custom-designed) & Shortcut paths in SBRSs & 2023 \\
    ~\cite{tsoumas2023evaluating} & Back-door adjustment, IPS &	(custom-designed, knowledge-based RS)& Digital agriculture & 2023 \\
    DDCE~\cite{wang2023dual}& - & (custom-designed) & Popularity bias & 2023 \\
    \bottomrule
    \end{tabular}}%
    \label{tab:confounder}%
    \begin{tablenotes}
    \small
    \item *Here, WOM stands for word-of-mouth, RL for reinforcement learning, and SBRS for session-based recommender system.
    \end{tablenotes}
    \vspace{-4mm}
\end{table*}%

\begin{figure}[t!]
\centering
\vspace{-1mm}
\includegraphics[height=0.13\textheight,trim=0 0 0 0 ,clip]{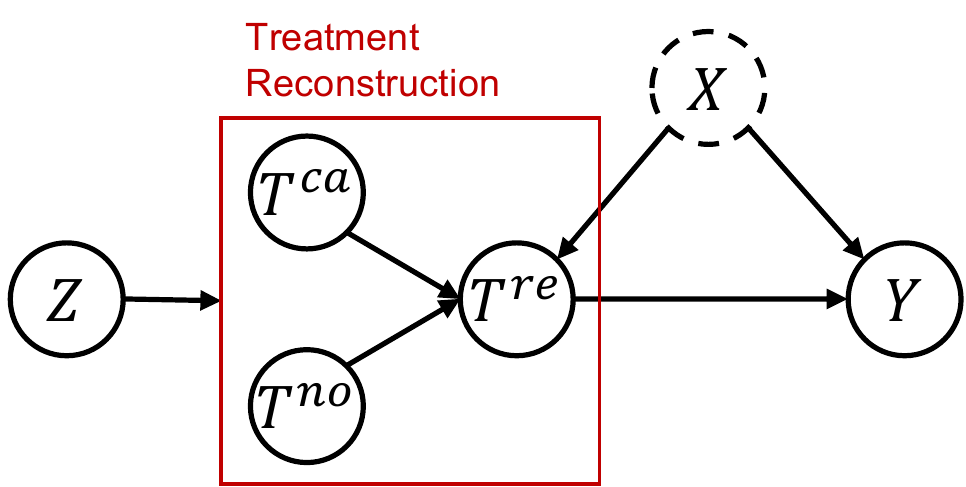} 
\vspace{-1mm}
\caption{The causal graph of IV4Rec, which reconstructs treatment $T$ by leveraging search queries $Z$ as instrumental variables to decompose treatment into causal part  $T^{ca}$ and non-causal part $T^{no}$ and combining them with different weights. $X$ is a set of unmeasurable confounders and $Y$ represents users’ interaction.}
\vspace{-1mm}
\label{fig:iv4rec}
\end{figure}

Though a popular tool, instrumental variable seems to find little application in recommender systems because of the difficulty of finding variables that satisfy the conditions of instrumental variables. As already cited above,  Sharma et al.~\cite{sharma2015estimating} utilize an instantaneous shock in direct traffic as an instrumental variable to evaluate the recommendation effect. Si et al.~\cite{si2022model}  propose a model-agnostic framework named IV4Rec that effectively decomposes the embedding vectors into two parts: the causal part indicating a user’s personal preference for an item, and the non-causal part merely reflects the statistical dependencies between users and items such as exposure mechanism and display position, with users’ search behaviors as the instrumental variable. More specifically, it modifies the traditional IV method, using the residual of the least square regression as the causal embedding instead of discarding it. The causal graph is illustrated in Fig. \ref{fig:iv4rec}.

Considering the stringent conditions often associated with IVs, a recent theoretical advancement~\cite{miao2023identifying} has been proposed to estimate treatment effects utilizing an auxiliary variable, which requires less restrictive prerequisites compared to IVs. An example causal diagram for auxiliary variables is visually represented in Fig. \ref{fig:iv_ex}(b), where $Z$ serves as a proxy variable for the unmeasurable confounder. Building on this theory, Zhang et al.\cite{zhang2023debiasing} developed the iDCF (identifiable deconfounder) to account for the unmeasured user’s socio-economic status $X$ by employing the user's consumption level as a proxy variable $Z$, a descendant of the unobserved confounder $X$ yet not directly causally associated with either treatment or outcomes. Furthermore, they leverage iVAE~\cite{khemakhem2020variational} to infer the conditional distribution of the latent confounder, thus resolving the Non-Identification issue encountered in ~\cite{wang2020causal}.

\subsubsection{The Front-door-based Approach}

The front-door adjustment~\cite{imbens2015causal} is another popular method for learning causal effects with unobserved confounders, in which we condition on a set of variables $K$ that satisfies the front-door criterion.

\begin{definition}[Front-door Criterion]
\label{def:frontdoor_criterion}
Given a pair of treatment $T$ and outcome variable $Y$, a set of variables $K$ is said to satisfy the front-door criterion if:
\begin{enumerate}
\item $K$ intercepts all directed paths from $T$ to $Y$;
\item there is no back-door path from $T$ to $K$; and
\item all back-door paths from $K$ to $Y$ are blocked by $T$.
\end{enumerate}
\end{definition}

\begin{figure}[t!]
	\centering
    \vspace{-3mm}
	\subfloat[Front-door path]{
		\centering
		\includegraphics[width=0.25\textwidth, trim=-10 -40 -10 0, clip]{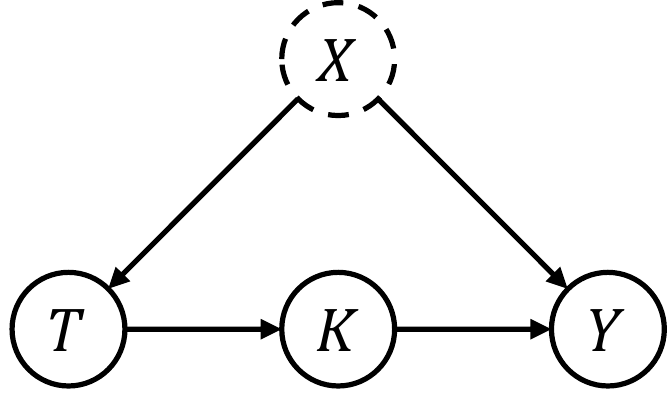}
	}
	\centering
	\subfloat[HCR]{
		\centering
		\includegraphics[width=0.25\textwidth, trim=-10 0 -10 0, clip]{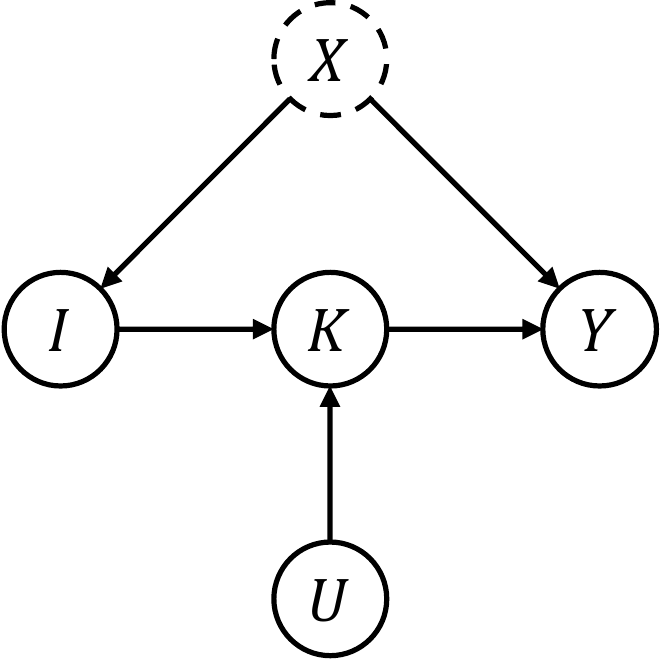}
	}
	\centering
    \vspace{-3mm}
	\caption{(a) A graphical model representing the front-door path, in which $T$ denotes the treatment, and $Y$ denotes outcomes. Unobserved confounders $X$ exist in the causal effect $T \rightarrow Y$, and $K$ are variables that satisfy the front-door criterion. (b) The Causal graph for illustrating the relationship in the HCR framework. $U$: user features, $X$: hidden confounders, $I$: item features affected by $X$, $Y$: post-click interactions, $M$: click behaviors. }
    \vspace{-6mm}
	\label{fig:frontdoor}
\end{figure}

A graph depicting the front-door criterion is shown in Fig. \ref{fig:frontdoor} (a). In practice, $K$ is usually the mediator of the causal effect $T \rightarrow Y$. With the help of $K$, the causal effect of $T$ on $Y$ can be calculated as follows:

\begin{definition}[Front-Door Adjustment)]
\label{def:frontdoor_adjustment}
If $K$ satisfies the front-door criterion relative to $(T, Y)$ and $\textrm{Pr}(T, Y) > 0$, then the causal effect of $T$ on $Y$ is given by the formula
\begin{equation}
\textrm{Pr}(Y \mid do(T)) =\sum_{K} \textrm{Pr}(Y \mid do(K)) \textrm{Pr}(K \mid do(T)) =\sum_{K} \textrm{Pr}(K \mid T) \sum_{T^{\prime}} \textrm{Pr}\left(Y \mid T^{\prime}, K\right) \textrm{Pr}\left(T^{\prime}\right).
\end{equation}
\end{definition}

Zhu et al.~\cite{zhu2022mitigating} propose HCR (Hidden Confounder Removal) framework to mitigate hidden confounding effects by front-door adjustment, in which user and item feature $U$ and $I$ are treatments, post-click user behaviors $Y$ are the concerned outcome, and the click feedback $K$ acts as the mediator that satisfies the front-door criterion, as Fig. \ref{fig:frontdoor} (b) shown. However, in real-world recommendation scenarios, confounding bias also exists in the estimation of the click feedback, which means it is not competent to perform the front-door adjustment. In fact, the front-door adjustment, like the IV method, finds little application in recommender systems because of the lack of eligible variables.

\subsubsection{Deconfounded Recommender Algorithms}

Instead of directly introducing causal technique, some literature expands sheer recommendation algorithms to deal with confounders under the inspiration of analysis from the perspective of causal inference. For example, ~\cite{chaney2018algorithmic} modifies several traditional recommendation algorithms to explore the impact of algorithmic confounding, which has found that the data-algorithm feedback loop amplifies the homogenization of user behavior without corresponding gains in utility and also amplifies the impact of recommendation systems on item consumption.

Some works integrate reinforcement learning-based recommender systems with causal inference to tackle the confounding issue. For example, DEMER (deconfounded multi-agent environment reconstruction)~\cite{shang2019environment} is proposed following the generative adversarial training framework to model the hidden confounder, which affects both actions and rewards as an agent interacts with the environment and thus obstructs an effective reconstruction of the environment, by treating the hidden confounder as a hidden policy. In ~\cite{yang2021top}, user representations $U$ are considered as a confounder of the recommendation lists $T$ and users’ interactions $Y$ on recommendation lists. To alleviate this confounding bias, CPR (counterfactual personalized ranking framework) builds the recommender simulator to generate new training samples based on the causal graph. 

As for session-based recommender systems (SBRSs), Gupta et al.~\cite{gupta2021causer} propose the CauSeR (Causal Session-based Recommendations) framework to perform deconfounded training to handle popularity bias. COCO-SBRS~\cite{song2023counterfactual} adopts a self-supervised approach to pre-train a recommendation model to learn the causalities in SBRSs, so as to eliminate confounding bias and make accurate next item recommendations. In terms of GNN-based recommendations, Gas et al. infer the unobserved confounders existing in representation learning with the CVAE model~\cite{sohn2015learning} and apply it to GNN-based strategy~\cite{gao2021deconfounding}.
\section{General Counterfactuals-based Methods}
\label{sec:general_counterfactual}
Some causal recommender approaches are established based on the general concept of counterfactuals, the world that does not exist but can be reasoned with some fundamental law and human intuition. In this section, we will introduce related strategies from the perspective of recommender issues they try to address, including domain adaptation, data augmentation, fairness, and explanation.

\subsection{Domain Adaptation}
\label{sec:domain_adaption}
RSs are trained and evaluated offline with the supervision of previously-collected data, which usually suffers from selection bias and confounding bias. It results in a gap between the training goal and the true recommendation objective, and, therefore, a sub-optimal recommender algorithm. To address this issue, we hope to evaluate the training policy on the unbiased data, which is collected from the randomized treatment policy. However, uniform data is always expensive and small-scale. To take full advantage of the uniform data, researchers train the recommender systems with a small amount of unbiased data and a large amount of biased data, with the hope of learning the counterfactual distribution of the biased data, which is both a counterfactual problem and a domain adaptation problem.

~\cite{rosenfeld2017predicting} and ~\cite{bonner2018causal} train recommender policies on biased and unbiased data, and add regulation terms to the loss function so that the distance of parameters between the two policies in the inspiration of individual treatment effect is controllable. ~\cite{yuan2019improving} trains an unbiased imputation model to impute the labels of all observed and unobserved events in biased and unbiased data, and learns the final CTR model by combining the two data with the propensity-free doubly robust method. Further, ~\cite{liu2020general} propose KDCRec (Knowledge Distillation framework for Counterfactual Recommendation) in which the teacher network with unbiased data as input is used to guide the biased model via four approaches.

\begin{table*}[pt]
  \small
  \centering
  \vspace{-3mm}
  \caption{\normalsize Summary of recommendation models with general counterfactuals.}
  \vspace{-3mm}
  \setlength{\tabcolsep}{1.9mm}{
    \begin{tabular}{c|c|c|c|c}
    \toprule
    \textbf{Category} & \textbf{Model} & \textbf{Causal inference method} & \textbf{Backbone model} & \textbf{Year} \\
    \midrule
    \multirow{4}[1]{*}{\makecell{Domain\\adaptation}} & ~\cite{rosenfeld2017predicting} & ITE   & Linear/regularized kernel methods & 2017 \\
          & ~\cite{bonner2018causal} & ITE   & Matrix factorization  & 2018 \\
          & Propensity-free DR~\cite{yuan2019improving} & DR    & FFM   & 2019 \\
          & KDCRec~\cite{liu2020general} & ITE   & MF (knowledge distillation) & 2020 \\
    \midrule
    \multirow{5}[0]{*}{\makecell{Data\\augmentation}} & CF2~\cite{xiong2021counterfactual} & "Minimum" counterfactuals & (custom-designed) & 2021 \\
          & CASR~\cite{wang2021counterfactual} & "Minimum" counterfactuals & NARM~\cite{li2017neural}, STAMP~\cite{liu2018stamp}, SASRec~\cite{kang2018self} & 2021 \\
          & CauseRec~\cite{zhang2021causerec} & Counterfactuals & \makecell{(custom-designed, sequential recommendation)} & 2021 \\
          & POEM~\cite{liu2022modeling} & Counterfactuals & GCN   & 2022 \\
          & COCO-SBRS~\cite{song2023counterfactual} & Counterfactuals & (custom-designed, sequential recommendation) & 2023 \\
    \midrule
    \multirow{4}[0]{*}{Fairness} & ~\cite{li2021towards} & Counterfactuals & (custom-designed) & 2021 \\
          & F-UCB~\cite{huang2022achieving} & Counterfactuals & UCB   & 2022 \\
          & CLOVER~\cite{wei2022comprehensive} & Counterfactuals & MELU~\cite{lee2019melu} & 2022 \\
          & PSF-RS~\cite{zhu2023path} & "Minimum" counterfactuals & (custom-designed) & 2023 \\
    \midrule
    \multirow{2}[0]{*}{Explanation} & PRINCE~\cite{ghazimatin2020prince} & "Minimum" counterfactuals & HIN~\cite{shi2016survey} & 2020 \\
          & CountER~\cite{tan2021counterfactual} & "Minimum" counterfactuals & MLP(black-box) & 2021 \\
          & CounterNet~\cite{guo2023counternet} & "Minimum" counterfactuals & (custom-designed) & 2023 \\
          \bottomrule
    \end{tabular}}%
  \label{tab:fair_explain}%
  \vspace{-6mm}
\end{table*}%

\subsection{Data Augmentation} 

Data augmentation is an uncontroversial counterfactual problem, such as answering the question: “what would be the user’s decision if a different item had been exposed?”. Therefore, some works are trying to integrate counterfactuals into the procedure of data augmentation.  

Xiong et al.~\cite{xiong2021counterfactual} generate new data samples by users’ feature-level preference for review-based recommendation. To generate more effective samples, they leverage the “minimum” idea in counterfactuals, learning the “minimum” change of the user feature-level preference that can “exactly” reverse the preference ranking of the user on a given item pair. For example, if slightly increasing the price attention of a user who had purchased an iPhone will make Xiaomi more attractive to her, this will be regarded as an effective counterfactual sample. Similarly, CASR (Counterfactual Data-Augmentation Sequential Recommendation)~\cite{wang2021counterfactual} generates the counterfactual sequence of items by “minimally” changing the user’s historical items, such that her currently interacted item can be “exactly” altered. 

The CauseRec (Counterfactual User Sequence Synthesis for Sequential Recommendation) proposed by Zhang et al.~\cite{zhang2021causerec} generates counterfactual data in a different way. It identifies indispensable and dispensable concepts in the historical behavior sequence. The former can represent a meaningful aspect of the user’s interest, while the latter indicates noisy behaviors that are less important in representing user interest. Therefore, it is reasonable to argue that replacing indispensable concepts in the original user sequence incurs a preference deviation of the original user representation, while replacing the dispensable ones still has a similar user representation, which CauseRec realizes through contrastive learning. Liu et al.~\cite{liu2022modeling} focus on the recommendation scenario where users are exposed with decision factor-based persuasion texts, i.e., persuasion factors, and generate new training samples by making simple but reasonable counterfactual assumptions about user behaviors, including:
\begin{itemize}
\vspace{-2mm}
    \item If a user clicks on an item without the existence of persuasion factors, the user will still be likely to click on it with a matching persuasion factor. 
    \item If a user does not click on an item with the existence of persuasion factors, the user will not click on it when the persuasion factor does not exist.
\vspace{-3mm}
\end{itemize}

In recent work, Song et al.~\cite{song2023counterfactual} categorize the factors influencing user interactions in session-based recommender systems into two types: inner-session causes and outer-session causes, and then generate counterfactual data samples through a novel combination of original inner-session causes and outer-session causes from similar users.

\subsection{Fairness and Explanation}

The counterfactual technique is a natural tool for the evaluation of fairness since we can compare the outcome (ratings, recommendation lists, etc.) in the real world and in the counterfactual world in which only users’ sensitive features (e.g., gender and race) are intervened~\cite{huang2022achieving, wei2022comprehensive}. 
\begin{definition}[Counterfactual Fairness]
\vspace{-3mm}
A recommender model is counterfactually fair if, for any possible user $u$ with features $X=x$ and $Z=z$:
\begin{equation}
\textrm{Pr}(y \mid x, z) = \textrm{Pr}(y \mid x, do(z’))
\end{equation}
For any value $y$ and $z’$, where $Y$ denotes the potential outcome for user $u$. $Z$ are users’ sensitive features and $X$ are causally $Z$-independent features.
\vspace{-3mm}
\end{definition}
Based on the counterfactual fairness, Li et al. generate sensitive feature-independent user embeddings through adversary learning~\cite{li2021towards}. They train a predictor to learn the filtered embedding and an adversarial classifier to predict the sensitive features from the learned representation simultaneously. For the reinforcement learning-based recommendation, Huang et al. propose the F-UCB (fair causal bandit)~\cite{huang2022achieving}, picking arms from a subset of arms at each round in which all the arms satisfy counterfactual fairness constraint that users receive similar rewards regardless of their sensitive attributes. Zhu et al.~\cite{zhu2023path} contend that directly removing or altering sensitive features will inevitably compromise the quality of recommendations, as these features can influence user interests fairly (e.g., racial influences on cultural preferences). To address this issue, their proposed PSF-RS (Path-Specific Fair Recommender System) delineates the influence process of sensitive features on interaction outcomes into fair and unfair paths, and addresses the path-specific bias by minimally transforming the biased factual world into a hypothetically fair one.

As for explanation, counterfactuals describe a dependency on the external facts that lead to certain outcomes, and thus allow researchers to reason about the behavior of a black-box algorithm~\cite{wachter2017counterfactual}. Literature on counterfactual explanation also resorts to the “minimum” idea in counterfactuals. For example, ~\cite{ghazimatin2020prince} presents PRINCE (Provider-side Interpretability with Counterfactual Evidence)  to search for a set of minimal actions performed by the user that, if removed, changes the recommendation to a different item, in a heterogeneous information network with users, items, and so on. To understand the point, consider the following example. If a user who has bought an iPhone and followed MacBook receives a recommendation about AirPods and would not have received it if she had not bought iPhone, PRINCE will regard the behavior “purchase of iPhone” as the explanation of the recommendation. Similarly, CountER (Counterfactual Explainable Recommendation) proposed by ~\cite{tan2021counterfactual} seeks the minimum changes of item features that exactly reverse the recommendation decision.

The aforementioned studies are predominantly post-hoc methods tailored for proprietary machine learning models, which restricts the explanatory models from leveraging information from the predictive models. The work by Guo et al.~\cite{guo2023counternet} introduces CounterNet, an integration that combines the predictive model and the counterfacutal explanation generator in an end-to-end framework. Beyond the scope of recommender systems, there are additional counterfactual explanation studies that may serve as supplementary references~\cite{wachter2017counterfactual, joshi2019towards, nemirovsky2022countergan,pawelczyk2020learning}.
\section{Opening Problems and Future Directions}
\label{sec:discuss}
 
The introduction of causal inference into recommender systems is still relatively recent, and there are still many promising but unexplored research directions, which we will discuss in this section.

\textbf{Causal assumptions in recommendations.} To extract causality knowledge from statistical data, causal inference-based methods are conducted with several causal assumptions. Although much of the existing work employ causal inference methods, these assumptions are often not explicitly clarified or even violated. To address this issue, existing work either ignores it or makes simple assumptions about data distributions. Therefore, estimating the impact of violations of causal assumptions on experimental results is crucial for bridging the gap between modern recommender system design and causal inference. For example, the positivity assumption is essential in PO-based approaches for the unbiased estimation of causal effect, but the data sparsity problem of recommender systems makes it difficult to satisfy. Considering that a small difference in recommendation accuracy may lead to a huge rise and fall in platform revenue, the effect of the violation of causal assumptions and that of artificial assumptions on predictions should be investigated.


\textbf{Causal Discovery in recommendations.} The integration of causal discovery and causal inference is inevitable in recommender systems. This belief stems from the fact that, as mentioned in Section \ref{sec:related}, the former serves as the foundation for the latter. In the absence of causal discovery, divergent causal graphs are often proposed for identical problems across different studies, leading to a plethora of methodologies, while overlooking the validation of the causal graph's correctness. Although these proposed methods have experimentally proven effective, they might lack generality and could be challenging to extrapolate to other datasets, a limitation notably prevalent in SCM-based methods. Causal discovery substantially reduces the reliance on manually designed causal graphs, enhancing the generalizability and applicability of causal recommendation methods across diverse scenarios. Possible research directions for Causal Discovery include: 1) Discovering causal relationships between variables. For example, exploring causal relationships between user attributes (e.g., age, economic status, and geographic context) and interaction decisions. 
2) Discovering causal relationships between users and items. For example, paper and ink cartridges are always simultaneously observed. But they are causally irrelevant; instead, they share a common cause - the item “printer”~\cite{wang2022sequential}. Accurately identifying item-level causal relationships significantly enhances the precision of recommendations. 
In this direction, some exploratory work has been done~\cite{wang2022sequential, he2022causpref}. One potential solution involves combining causal discovery with GNNs. The ability of GNNs to explore structural information between nodes in a graph~\cite{gao2022graph} gives them a natural advantage in identifying causal relationships between users and items. Furthermore, the causal knowledge uncovered can be further incorporated into GNN-based recommendation algorithms to facilitate the learning of semantically meaningful and identifiable graph representations~\cite{jiang2023survey}.

\textbf{Transfer learning and Out-of-distribution recommendation.} Due to the data sparsity issue of recommendation systems, it will be a wise and practical choice to transfer user and item knowledge from other domains to improve prediction performance during cold start, offline evaluation, or online test, which is a transfer learning problem. Even in the same dataset, natural shifts of user preference or artificial bias also cause a violation of the IID hypothesis, which is an out-of-distribution (OOD) recommendation issue ~\cite{he2022causpref}. The core of both transfer learning and OOD recommendation is to transfer beneficial shared knowledge, such as users' inherent and unchanged preferences. Thus they can be formulated as invariant learning in some cases~\cite{wang2022invariant}. As we mentioned in Section \ref{sec:intro}, causal inference works to discover the unchangeable causal relationship in data, which can be reused in new domains. From this perspective, adopting causal inference to improve robustness and generalization ability to accomplish cross-domain or OOD recommendation is a promising direction, and some exciting attempts can be found recently~\cite{he2022causpref, wang2022out}.

\textbf{Dynamic recommendation.} Modern recommender systems usually involve feedback loops and dynamic updates. Therefore, it is crucial to incorporate loops into the causality-based methods to accurately model the dynamic and iterative data collection process for recommender systems~\cite{xu2022dynamic}. Some impressive work has also been proposed~\cite{chaney2018algorithmic, wang2021deconfounded, gupta2021causer, krauth2022breaking}. However, uncontrolled feedback loops may give rise to issues like the Matthew effect, echo chambers~\cite{chaney2018algorithmic, ge2020understanding, xu2022dynamic} and bias amplification~\cite{chaney2018algorithmic, wang2021deconfounded}. Original debiasing approaches (e.g., back-door adjustment) cannot be applied directly due to the change in the form of causal models. Therefore, deconfounding in multi-step and feedback loop-involved causal models is still an open research field.

\textbf{Causality-inspired foundation models for recommendation.}
The emergence of Large Language Models (LLMs) has sparked extensive exploration into the development of recommendation foundation models. These models, pre-trained on diverse language or interaction data, can be adapted for a wide array of downstream recommendation tasks~\cite{liu2023first, qiu2021u, hou2022towards, kang2023llms, lin2023can}. Integrating causality into these models is a promising direction~\cite{petrov2023generative}. However, it has been observed that bias in the pretraining corpus of foundation models can lead to unfairness in recommender systems from both user-side~\cite{hua2023up5, zhang2023chatgpt} and item-side~\cite{hou2023large}. Current studies primarily focus on the fairness issue in specific recommendation tasks. To address this, there is a growing interest in formulating novel pretraining tasks to evaluate the causal inference capabilities of recommendation foundation models~\cite{jin2023can}, aiming to mitigate bias issues at their root and enhance overall recommendation performance.

\section{Conclusion}

In this survey paper, we have summarized the mechanisms and the strategies of causal inference for recommender systems, from the theoretical perspective: PO framework-based, SCM framework-based and general counterfactuals-based. The survey gives the clear description about the strengths of causal inference for recommendations and manages to use a uniﬁed symbol system to describe a large number of existing causal recommender approaches. We hope this survey can well help researchers in the recommendation field to utilize and innovate.

\begin{acks}
This research work is supported by the National Key Research and Development Program of China under Grant No. 2021ZD0113602, the National Natural Science Foundation of China under Grant Nos. 62176014, 62276015, the Fundamental Research Funds for the Central Universities.
\end{acks}

\section*{Declaration of Interests}

The authors declare no competing interests that could have appeared to influence the work reported in this paper.

\bibliographystyle{ACM-Reference-Format}
\bibliography{ref}





\end{document}